\documentclass[aps,pre,reprint,twocolumn,showpacs,superscriptaddress,longbibliography]{revtex4-2}
\usepackage{graphicx}
\usepackage{dcolumn}
\usepackage{bm}
\usepackage{hyperref}
\usepackage{amsmath}
\usepackage{diagbox}

\usepackage{longtable}
\usepackage{color}
\usepackage{xcolor}

\begin{document}

\title{Effect of substrate heterogeneity and topology on epithelial tissue growth dynamics}

\author{Mahmood Mazarei}
\email[]{mmazare@uwo.ca}
\affiliation{Department of Physics and Astronomy,
  Western University, 1151 Richmond Street, London,
  Ontario, Canada  N6A\,3K7}

\author{Jan {\AA}str{\"o}m}
\email[]{jan.astrom@csc.fi}
\affiliation{CSC Scientific Computing Ltd, K{\"a}gelstranden 14, 02150 Esbo, Finland}

\author{Jan Westerholm}
\email[]{jan.westerholm@abo.fi}
\affiliation{Faculty of Science and Engineering, {\AA}bo Akademi
University, Vattenborgsv\"agen 3, FI-20500, {\AA}bo, Finland
}

\author{Mikko Karttunen}
\email[]{mkarttu@uwo.ca}
\affiliation{Department of Physics and Astronomy,
  Western University, 1151 Richmond Street, London,
  Ontario, Canada  N6A\,3K7}
\affiliation{Department of Chemistry,  Western University, 1151 Richmond Street,
London, Ontario, Canada N6A\,5B7
}

\date{\today}

\begin{abstract}
Tissue growth kinetics and interface dynamics depend on the properties of the tissue environment and cell-cell interactions. In cellular environments, substrate heterogeneity and geometry arise from a variety factors, such as the structure of the extracellular matrix and nutrient concentration. We used the \textsc{CellSim3D} model, a kinetic division simulator, to investigate the growth kinetics and interface roughness dynamics of epithelial tissue growth on heterogeneous substrates with varying topologies. The results show that the presence of quenched disorder has a clear effect on the colony morphology and the roughness scaling of the interface in the moving interface regime. In a medium with quenched disorder, the tissue interface has a smaller interface roughness exponent, $\alpha$, and a larger growth exponent, $\beta$. The scaling exponents  also depend on the topology of the substrate and cannot be categorized by well-known universality classes.
\end{abstract}
\maketitle

\section{Introduction}

Understanding the role of mechanobiological phenomena in  complex biological processes such as wound healing, tumor growth, and morphogenesis necessitates the study of the physical interactions between cells and their environments. \textit{In vivo}, heterogeneities of different types are always present. One of the prime examples is the extracellular matrix (ECM) that typically provides support for cells and is a key factor for cell adhesion and the differentiation of cells~\cite{kleinman2003role,frantz2010extracellular}. Heterogeneities can also be produced by the addition of pharmacological agents or (gelly) materials, such as methylcellulose, or by changing the nutrient concentration, as well as by other means~\cite{Huergo2014,rapin2021roughness,galeano2003dynamical}. The presence of heterogeneities, or disorder in physical terms, often influences biochemical and biomechanical parameters, such as cell-cell interactions, the rate of cell division, and the average cell size and shape, and thus alters cell mobility, colony spreading, and the roughness of the colony interface~\cite{Huergo2014,rapin2021roughness}. In addition, the situation can be even more complex such as in the epithelial-to-mesenchymal and mesenchymal-to-epithelial transitions during which the whole cellular environment undergoes fundamental and complex changes~\cite{Pastushenko2018-ak}.

Tumor growth, and therapy to prevent it, may be characterized as cellular processes involving molecular inter- and intracellular control~\cite{sengupta2021principles}. Cell migration continually responds to the mechanical stresses from 
neighboring cells and the ECM~\cite{Ventura2022-ev}. Mathematical and computer models are increasingly being used to examine and measure the influence of different biophysical parameters on biological processes such as nonequilibrium pattern generation in biological growth~\cite{buttenschon2020bridging,wortel2021artistoo,conradin2021palacell2d,li2021role,rapin2021roughness}.

The spreading of a cellular colony, e.g., tumor, healthy or bacterial, can be seen as the propagation of an elastic interface in the presence of a pinning potential that arises from the surrounding enviroment. Analogous phenomena occur in diverse systems including vortex motion in type-II superconductors~\cite{larkin1979pinning}, charge-density waves~\cite{balents1995temporal,Karttunen1999-ta}, and fracture propagation~\cite{bouchaud1993models}. For such systems, one typically distinguishes between strong and weak pinning. 
In the former case, the pinning energy (per impurity) is much larger than the elastic energy wleading to local energy minimization while in the weak pinning regime the opposite is true and the interface adjusts collectively. 

When the interface has adjusted to the disorder and is not moving, it is in the pinned phase. When a driving force is applied and it exceeds a threshold force, $F_{c}$, the interface undergoes a depinning transition and enters the moving phase. The size of the advancing regions is then characterised by a correlation length ($\xi$) which diverges upon approaching the critical force from above, $\xi = \left ( F - F_{c} \right )^{- \nu}$, where $\nu$ denotes the correlation length exponent~\cite{Fisher1983-tz}. It is also common to differentiate between annealed and quenched disorder. In the latter, disorder is considered as stationary, that is, the motions of the pinning sites are much slower than any other relevant time scale in the system; in the annealed case this assumption no longer holds. In this study, only quenched disorder is considered. In addition, since here the driving force enters through cell division, we are not interested in the depinning transition itself.

Several dynamic universality classes have been proposed for interface growth. The Kardar-Parisi-Zhang (KPZ) dynamic universality class~\cite{Kardar1986} describes the evolution of a surface using a continuous nonlinear stochastic differential equation
\begin{equation}
\label{eq:kpz}
\partial_{t}h(x,t) = -\lambda \left[\partial_{x} h(x,t)\right]^{2} + \nu \partial_{x}^{2} h(x,t) + \eta(x,t), 
\end{equation} 
where $h(x,t)$, or the height, is the distance from
the $i$th point at the colony front to the
baseline
of the colony. Lateral growth normal to the interface, reflected in the quadratic term $-\lambda (\partial_{x} h)^{2}$,  is a characteristic of the KPZ universality class. Surface tension is accounted for by the Laplacian term, $\nu \partial_{x}^{2} h$, which tends to flatten the surface, and $\eta(x,t)$ is an uncorrelated Gaussian noise given
by $\langle \eta(x,t) \rangle=0$ and $\langle \eta(x,t) \eta(x',t')\rangle = 2 D \delta(x-x') \delta(t-t')$. 

In the quenched KPZ (qKPZ) equation the thermal noise in Eq.~\ref{eq:kpz}
is replaced by a position dependent noise, that is, 
$\eta(x,t)$ becomes $\eta(x,h)$ with
$\langle \eta(x,h) \rangle=0$ and $\langle \eta(x,h) \eta(x',h')\rangle = 2 D \delta(x-x') \delta(h-h')$. Since thermal noise is usually present in experiments, the qKPZ equation can be extended to contain both quenched and thermal noise.

Dynamic scaling analysis provides powerful tools to classify growth. Dynamic critical exponents, namely the roughness ($\alpha$), growth ($\beta$), and the dynamic exponent ($z$) can be determined from the time evolution of the front's roughness; the cell colony's  front in this case~\cite{costa2015universal,li2021role,khain2021dynamics,radszuweit2009comparing,Bru2003,Bru1998,Bru2005,Huergo2010,Huergo2011}. In addition to the above, the dynamic exponent is related to the two other 
exponents via $z = \frac{\alpha}{\beta}$.

Distinct critical exponents and universality classes are described by different growth equations. For KPZ, the critical exponents are $\alpha^\mathrm{KPZ} \!= \! \frac{1}{2}$, $\beta^\mathrm{KPZ} \!= \! \frac{1}{3}$, and $z^\mathrm{KPZ} \!= \! \frac{3}{2}$ for one dimensional interfaces~\cite{Kardar1986}. For the quenched KPZ equation, dynamic critical exponents haven been determined to be $\alpha^\mathrm{qKPZ} \!= \! \frac{3}{4}$, $\beta^\mathrm{qKPZ} \!= \! \frac{3}{5}$, and $z^\mathrm{qKPZ} \!= \! \frac{5}{4}$~\cite{csahok1993dynamics}. The critical exponents of the linear molecular beam epitaxy (MBE) equation for a one-dimensional interface are $\alpha^\mathrm{MBE} \!= \! \frac{3}{2}$, $\beta^\mathrm{MBE} \!= \! \frac{3}{8}$, and $z^\mathrm{MBE} \!= \! 4.0$.

In one-dimensional quasilinear and quasicircular expanding interfaces, previous experimental research on cells grown on culture without quenched disorder have presented various scaling behaviors~\cite{Bru1998,Bru2003,Huergo2010,Huergo2011,Huergo2012}. Br\'u \textit{et al.}~\cite{Bru2003,Bru1998} suggested that the development dynamics of both malignant and normal cell colonies 
are characterized exponents $\alpha = 1.5 \pm 0.15$, $\beta = 0.38 \pm 0.07$, and $z = 4 \pm 0.5$ that belong to the MBE universality class. 
The reported this for both \textit{in vitro} and \textit{in vivo} experiments. In contrast, however, Huergo \textit{et al.} reported the exponents of $\alpha=0.50 \pm 0.05$, $\beta=0.32 \pm 0.04$ and $z = 1.56 \pm 0.1$ for interfacial growth of HeLa (cervix cancer) cell colonies \textit{in vitro}~\cite{Huergo2010,Huergo2011,Huergo2012}. Plant calli, \textit{Brassica oleracea} and \textit{Brassica rapa}, were studied by Galeano \textit{et al.} who reported exponents inconsistent with both MBE and KPZ,  $\alpha = 0.86 \pm 0.04$, and $z = 5.0$~\cite{galeano2003dynamical}. 

Biological systems with substrate disorder appear in situations such as growing bacterial colonies on agar-containing media and in the development of bacterial biofilms. For \textit{Escherichia coli} and \textit{Bacillus subtilis} colonies, Vicsek \textit{et al.}~\cite{Vicsek1990} found the roughness exponent $\alpha\! =\! 0.78\! \pm \!0.07$, which exceeds the KPZ value. 
Huergo \textit{et al.} reported qKPZ-compatible exponents $\alpha \!= \! 0.63 \pm  0.04$, $\beta \!= \! 0.75 \pm 0.05$, and $z \!=\! 0.84 \pm 0.05$ for the development of quasilinear Vero cell colony fronts in culture media containing methylcellulose (MC)~\cite{Huergo2014}. Santalla \textit{et al.} conducted experiments at a high agar concentration regime and found branching interfaces whose scaling exponents were in complete disagreement with both the KPZ and qKPZ scaling exponents~\cite{santalla2018nonuniversality}. Rapin \textit{et al.} studied the effects of pharmacological agents on the geometry and roughness dynamics of \textit{in vitro} propagating Rat1 fibroblast cell interfaces and reported two separate scaling regimes,  the first at the sub-cell level and the second at intermediate length scales of 2-10 cells~\cite{rapin2021roughness}.

Various theoretical and computational models have been developed to examine surface growth with quenched disorder. The directed percolation depinning model predicts $\alpha$ to be between $0.66$ and $0.73$, and $\beta \!=\! 0.68 \pm 0.04$~\cite{Buldyrev1992}. Models of self-organized growth have predicted $\beta \!=\! 0.9 \pm 0.1$ and $\alpha \!=\! 0.63 \pm 0.02$~\cite{Sneppen1992}, and a numerical study of an automaton model yielded $\alpha \!=\! 0.63 \pm 0.01$ and $\beta \!=\! 0.64 \pm 0.02$~\cite{Leschhorn1996}. Santalla and Ferreira incorporated nutrient diffusion to an off-lattice Eden model and reported a transition from a transient KPZ-like regime with $\beta \!=\! 0.34 \pm 0.01$ to an unstable growth regime with $\beta \!=\! 0.43 \pm 0.02$, with an intermediate transient regime belonging to the qKPZ universality class with $\beta \!=\! 0.633$ and the local roughness exponents ranging within $0.39 < \alpha_{\mathrm{loc}} < 0.67$~\cite{santalla2018eden}. 

Further computational and theoretical studies have demonstrated the effects of cell-cell mechanical tensions, and nutrient concentration and distribution on the spatial structures with morphologies ranging from smooth to heavily fingered interfaces~\cite{Young2022-ne,Wang2017-yx}.
Simulations of two-dimensional cellular colonies by Block \textit{et al.}, showed KPZ-like dynamics for a class of cellular automata models over a broad range of parameters~\cite{Block2007}. Azimzade \textit{et al.} used the Fisher-Kolmogorov-Petrovsky-Piskunov (FKPP) equation to study the effect of the cellular environment's stiffness and spatial correlations on the morphology of the interface of growing tumors, and concluded that the KPZ  equation cannot describe their tumor development model~\cite{azimzade2019effect}. 
Bonachela \textit{et al.} developed an off-lattice cell model with quenched disorder  
describing competition among bacterial cells for space and resources. They reported the exponents $\alpha \,=\, 0.68 \pm 0.05$, $\beta \,=\, 0.61 \pm 0.05$, and $z \,=\, 1.11 \pm 17$ for the moving regime~\cite{Bonachela2011-wj}. 
Pinto \textit{et al.} modified the self-propelled Voronoi model of Bi \textit{et al.}~\cite{bi2016motility} to study the effect of spatial disorder of the cell-substrate interaction, defined as having stiff cells in the tissue, on cell motility in a confluent tissue, reporting $\beta \,=\, 0.194. \pm 007$~\cite{Pinto2022-ak}. In our previous work, we showed that a cell colony can show both KPZ- and MBE-like scaling dynamics depending on the strength of the cell-cell adhesion between the cells and the cell colony's geometry~\cite{mazarei2022silico}.

\section{Methods}

\subsection{\textsc{CellSim3D simulator and model}}

\textsc{CellSim3D} is a coarse-grained molecular dynamics\-based model of cellular dynamics with an emphasis on mechanobiological features of tissue growth~\cite{Madhikar2018}.  The code is open source~\cite{cellsim3d-web}. 
\textsc{CellSim3D} allows cellular growth to be 
modelled in two (epithelial growth) or three dimensions, and cells are modelled as three-dimensional objects consisting of a set of interconnected nodes. Here, the geometry and the nodes are those of  a spherical C180 fullerene. 

The \textsc{CellSim3D} force field consists of intra- and intercellular forces and a noise term ($\eta$),
\begin{equation}
\label{eq:Force}
m \mathbf{ \ddot{r} } = \mathbf{F}^{\mathrm{B}} + \mathbf{F}^{\mathrm{\theta}} + \mathbf{F}^{\mathrm{R}} + \mathbf{F}^{\mathrm{A}} + \mathbf{F}^{\mathrm{F,e}} + \mathbf{F}^{\mathrm{F,m}} + \mathbf{F}^{\mathrm{P}} + \eta .
\end{equation} 
The two intracellular forces on the surfaces of the cells are $\mathbf{F}^\mathrm{B}$, a damped harmonic oscillator force between the nearest neighboring nodes with a spring constant ($k^\mathrm{B}$) and a friction coefficient ($\gamma_\mathrm{int}$), and $\mathbf{F}^\mathrm{\theta}$, the angle force which is a harmonic potential depending on the equilibrium angles between the nodes with a spring constant $k^\mathrm{\theta}$). The angle term preserves the cell's surface curvature. For simplicity, the spring constants for both the angle force and the damped spring force between the nodes are assumed to be constant over the cell surface. 

Intercellular forces in \textsc{CellSim3D} consist of both cell-cell and cell-environment interactions. In cells, the cell-cell interactions are mainly caused by cell adhesion molecules (CAMs)~\cite{Murray1999,Edelman1991,Stewart2011}. Here, the intercellular forces are described by a repulsive force, $\mathbf{F}^\mathrm{R}$ and an attractive force, $\mathbf{F}^\mathrm{A}$, between two neighboring cells. In addition, the model also includes a friction force, $\mathbf{F}^\mathrm{F}$, between two cells that pass by each other. The repulsive and attractive forces between the cells are represented, respectively, by short-range harmonic potentials with distinct cutoffs $R_{0}^\mathrm{R}$, $R_{0}^\mathrm{A}$, and spring constants $k^\mathrm{A}$, $k^\mathrm{R}$. In this study, we assume that the adhesion molecules are distributed uniformly across the cell surface, and that the adhesion and repulsion spring constants ($k^\mathrm{A}$, $k^\mathrm{R}$) are identical for all nodes on the surface. The intermembrane friction force, $\mathbf{F}^\mathrm{F,e} = - \gamma_\mathrm{ext} \, \mathbf{v}^\mathrm{\tau_{m}}_{ij}$, is defined up to a cutoff range, $R^{\mathrm{A}}_{0}$, between the nodes $i$ and $j$ on two separate cells as a function of the tangential relative velocity to the cell surfaces, $\mathbf{v}^\mathrm{\tau_{m}}_{ij}$. The intermembrane friction coefficient, $\gamma_\mathrm{ext}$, is assumed to be constant across the cells.

The friction force, $\mathbf{F}^\mathrm{F,m} = - \gamma_\mathrm{m} \, \mathbf{v}$, approximates the interactions between the cell and its environment, and it is defined as a viscous drag force from a fluid medium. The growth force, $\mathbf{F}^\mathrm{P} = PS\mathbf{\hat{n}}$, is determined by the cell's internal pressure resulting from the osmotic pressure within the cell~\cite{Murray1999}, where $\mathbf{\hat{n}}$ is 
an outward pointing normal to the surface of the cell and $PS$ is the force due to 
a growing pressure inside the cell. This growing force compensates for the cell membrane elasticity modelled by harmonic potentials. Finally the noise term, $\eta$, is defined as a Gaussian white noise with $\langle \eta(x,t) \rangle=0$ and $\langle \eta(x,t) \eta(x',t')\rangle \! = \! 2 D \delta(t-t') \delta(x-x')$. 

At each time step, the internal pressure increases 
by the growth rate $\Delta (PS)$, resulting in a gradual increase in the pressure force ($\mathbf{F}^\mathrm{P}$) and the cell volume. When the volume of the cell reaches a critical threshold, given by the parameter $V_\mathrm{div}$, the cell divides into two daughter cells. The distinguishing characteristics of the cell division are the orientation and the location of the division plane. Cell division can be either symmetric or asymmetric, depending on the position of the division plan. In this study, we used symmetric cell division, in which the volumes of the daughter cells become half the volume of the parent cell, and the mechanical properties are a copy of the parent cell's properties. The division algorithm is accounts for the planar expansion of epithelial tissue: The division plane is selected by randomly sampling a vector from a circle in the plane defined by the vector normal to the epithelial plane. To prevent buckling during growth, three-dimensional cells are confined between two frictionless plates with repulsion in the direction normal to the plates~\cite{Madhikar2020,mazarei2022silico}. More details of the theoretical basis, the code implementation, and the mapping of the parameters can be found in Refs.~\cite{Madhikar2018,mazarei2022silico,Madhikar2020}. Parameters for the simulations performed in this study are provided in Table~\ref{tab:SimUnits}.
\begin{table}[h!]
  \begin{center}
    \caption{The parameters for the cells used in this study. These values are based on the HeLa (named after Henrietta Lacks~\cite{Scherer1953-hp}) cell properties. $\dagger$ indicates units of $\Delta t$ and $\ast$ units of mean time to cell division, which varies between cell types and is set to 1.0 in \textsc{CellSim3D}.}
    \label{tab:SimUnits}
	\resizebox{1\columnwidth}{!}
	{\tiny \begin{tabular}{l c c c c}
      \hline
      Parameter & Notation & Sim. Units & \multicolumn{2}{c}{SI Units} \\
      \hline
      Nodes per cell  & $N_\mathrm{c}$ & 180 & - & \\
	  Node mass  & m & 0.04 & 40 & fg\\
	  Bond stiffness  & $k^\mathrm{B}$ & 1000 & 100 & $ \mathrm{nN}/ \mu \mathrm{m}$\\
	  Bond damping coefficient  & $\gamma_\mathrm{int}$ & 100 & 0.01 & $\mathrm{g} /\mathrm{s}$\\
	  Minimum pressure  & $(PS)_{0}$ & 50 & 0.5 & $\mathrm{nN} / \mu \mathrm{m}^{2}$\\
	  Maximum pressure  & $(PS)_\mathrm{\infty}$ & 65 & 0.65 & $\mathrm{nN} / \mu \mathrm{m}^{2}$\\
	  Pressure growth rate  & $\Delta(PS)$ & 0.002 & $2.0 \times 10^{-5}$ & $\mathrm{nN} / \mu \mathrm{m}^{2}$\\
	  Attraction stiffness  & $K^\mathrm{A}$ & 10-2000 & 1-200 & $\mathrm{nN} / \mu \mathrm{m}$\\
   	Strong attraction stiffness  & $K^\mathrm{A}_\mathrm{strong}$ & 2000 & 200 & $\mathrm{nN} / \mu \mathrm{m}$\\
    Weak attraction stiffness  & $K^\mathrm{A}_\mathrm{weak}$ & 10 & 1 & $\mathrm{nN} / \mu \mathrm{m}$\\
	  Attraction range  & $R^\mathrm{A}_{0}$ & 0.3 & 3 & $\mu \mathrm{m}$\\
	  Repulsion stiffness  & $K^\mathrm{R}$ & $ 10 \times 10^{5}$ & $10 \times 10^{4}$ & $\mathrm{nN} / \mathrm{m}$ \\
	  Repulsion range  & $R^\mathrm{A}_{0}$ & 0.2 & 2 & $\mu \mathrm{m}$\\
	  Growth count interval  & - & 1000 & $\dagger$ & \\
	  Inter-membrane friction  & $\gamma_\mathrm{ext}$ & 1 & 10 & $\mu \mathrm{g} / {s}$\\
	  Medium friction  & $\gamma_\mathrm{m}$ & 0.4 & 4 & $\mu \mathrm{g} / s$\\
	  Time step  & $\Delta t$ & $1.0 \times 10^{-4}$ & $\ast$ & \\
	  Threshold division volume  & $V^\mathrm{div}$ & 2.9 & 2900 & $\mu \mathrm{m}^{3}$\\
    \end{tabular}}
  \end{center}
\end{table}

\subsection{Disorder}

Pinning impurities
were randomly positioned (at time $t=0$) as immobile cells that do not grow. They interact with regular cells via adhesion, repulsion, and friction, with the same strengths as the regular cells do. 
Importantly, when the adhesion interaction between the cells is strong, so is the interaction between the cells and the disorder. The same applies for the case of weak cell-cell interaction. The pinned cells maintain their spherical shapes and sizes throughout the simulation. 
For each simulated parameter set, ten independent simulations were performed for data averaging. The parameters for quenched disorder are shown in Table~\ref{tab:DisorderDens}. 
\begin{table}[!htbp]
    \begin{center}
    \caption{The area density, $\frac{N}{A}$, for quenched disorder in SI units ($\frac{1}{\mu \mathrm{m}^{2}}$) in the different configurations (linear and radial), and  at different attraction stiffnesses ($\frac{\mathrm{nN}}{\mu \mathrm{m}}$). The parameters for the strong and weak cases are given in Table~\ref{tab:SimUnits}.}
    \label{tab:DisorderDens}
    \resizebox{1\columnwidth}{!}
    {\tiny \begin{tabular}{ | l | c | c |}
    \hline
    \diagbox[width=15em,height = 2em]{Configuration}{Attraction stiffness}  &  1 (weak) & 200 (strong) \\
    \hline
    Moving linear interface                    & 0.0010    & 0.0003   \\
    \hline
    Moving linear interface at high disorder density                   & 0.0012    & -   \\
    \hline
    Moving radial interface                   & 0.0008     & 0.000075  \\
    \hline
    Pinned radial interface                   & 0.0012     &   -   \\ 
    \hline
    \end{tabular}}
    \end{center}
\end{table}

\begin{figure*}
\resizebox{.43\columnwidth}{!}{ \includegraphics{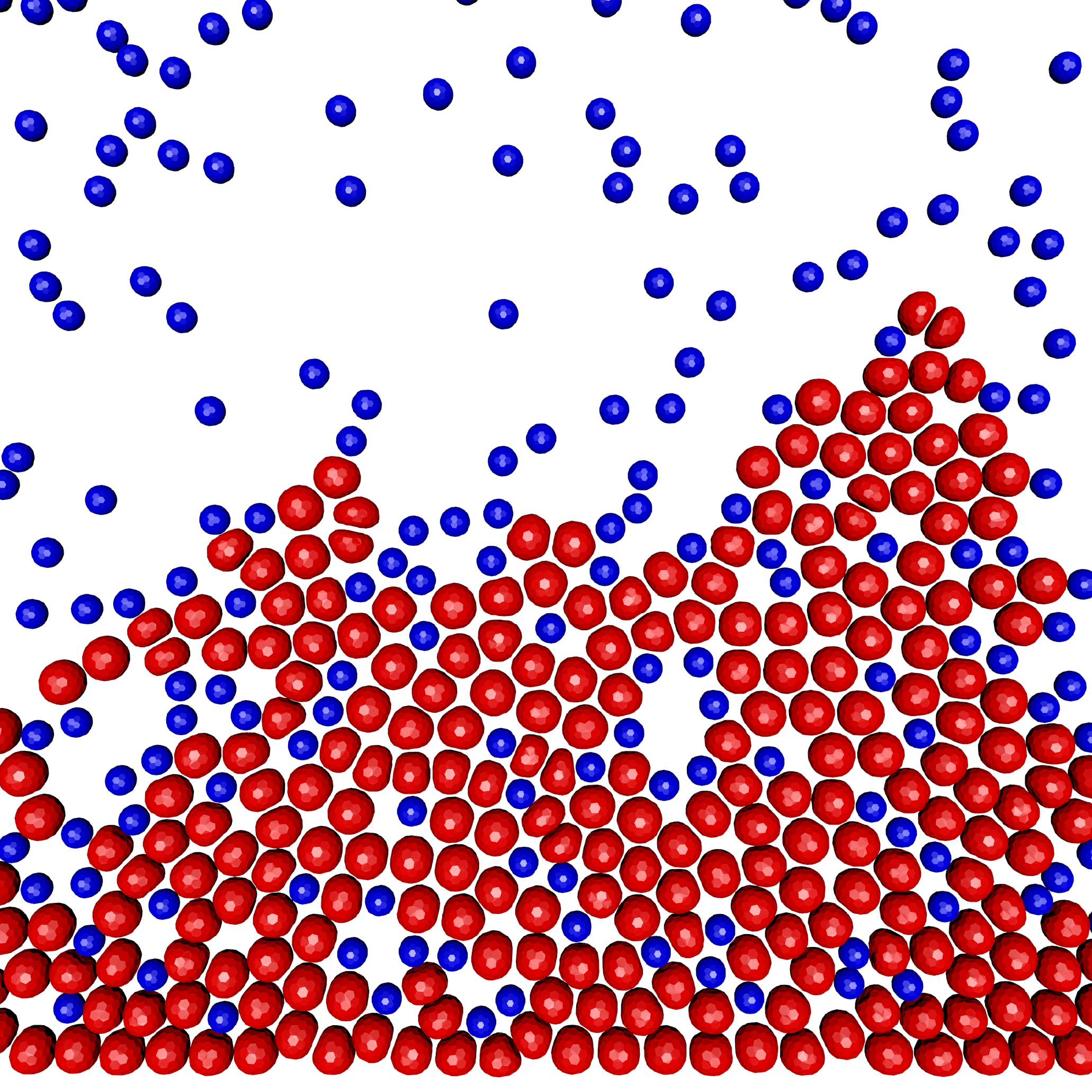} }
\hspace{22pt}\resizebox{.43\columnwidth}{!}{ \includegraphics{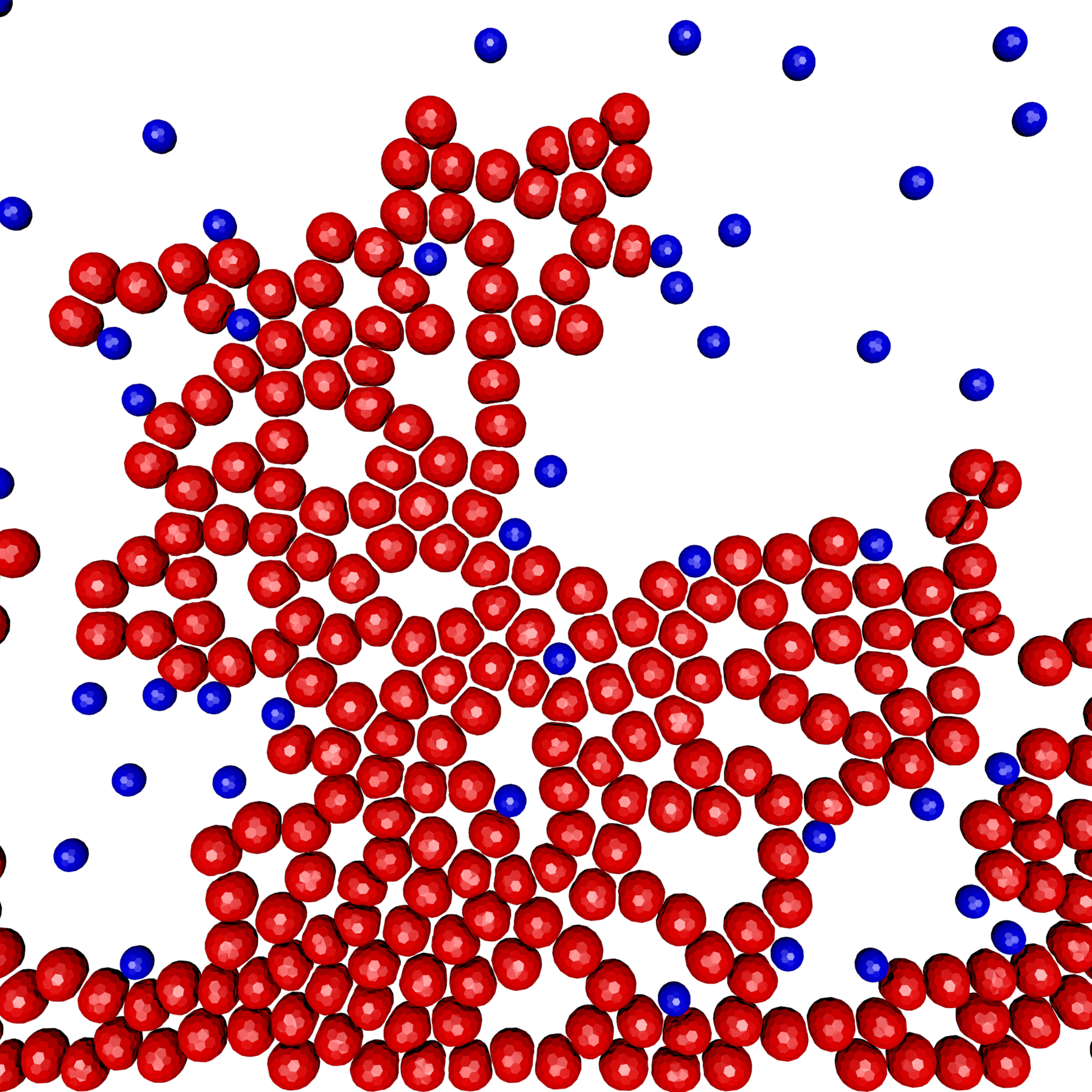} }
\hspace{12pt}\resizebox{0.51\columnwidth}{!}{ \includegraphics{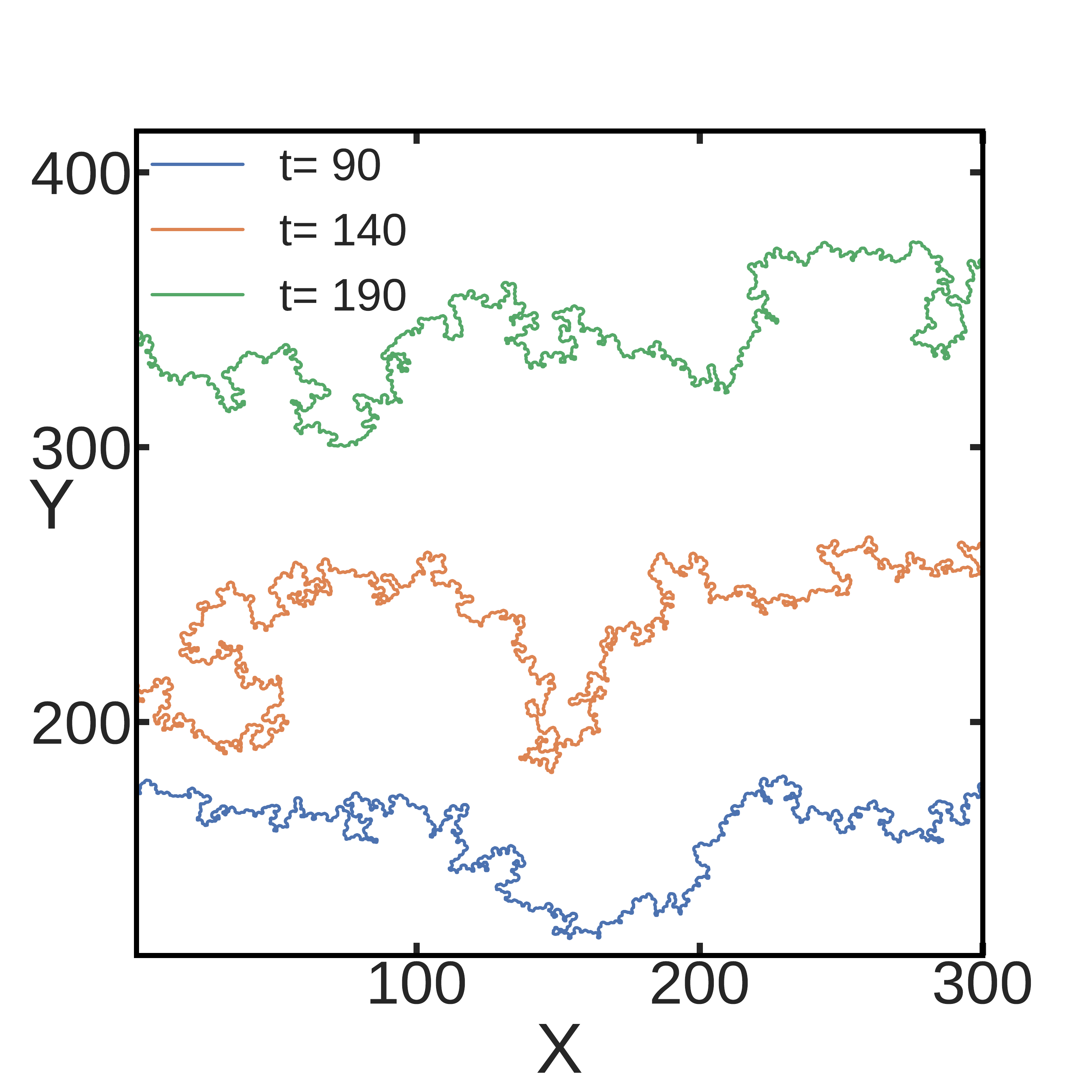} }
\resizebox{0.51\columnwidth}{!}{ \includegraphics{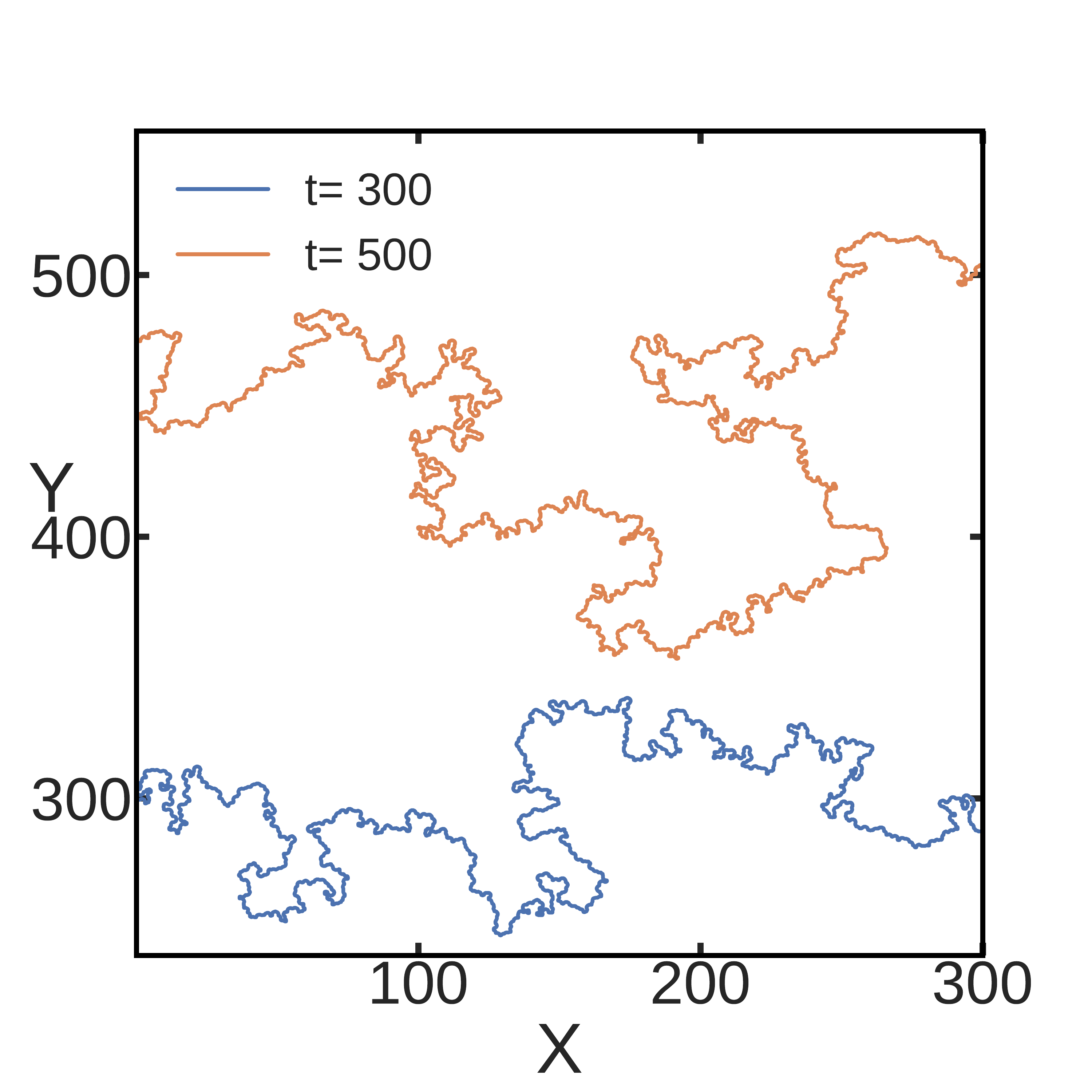} }\\
a) \hspace{0.5\columnwidth} b) \hspace{0.5\columnwidth} c) \hspace{0.5\columnwidth} d) 
\caption{(a-b) Colony expansion (red cells) in a medium with quenched disorder (blue cells) with linear initial configuration. (a) At weak and (b) strong cell-cell adhesion strength. (c-d)  Interface evolution at different times (c) at weak and (d) strong cell-cell adhesion strength. The scaling analysis was done using overhang-corrected interfaces~\cite{Barabasi1995}. For units, see Table~\ref{tab:SimUnits}.}
\label{fig:LineConfig}
\end{figure*}

\subsection{Colony configurations}

Simulations of both linear and radial growth at strong and weak cell-cell adhesion strengths in the presence of quenched disorder were performed at both low and high disorder densities, see Table~\ref{tab:SimUnits} for parameters and Table~\ref{tab:DisorderDens} for disorder area densities. The initial configuration of the linear interface was a line of 240 cells in a box of size $600\times 1,000\times 1.8$. For linear interfaces in the low disorder density regime at weak and strong cell-cell adhesion, 60,000 and 18,000 immobile cells were initially distributed at random in the box, while in the high disorder density regime at weak cell-cell adhesion, 72,000 immobile cells were randomly distributed in the box, see Table~\ref{tab:DisorderDens} for disorder area density. Figure~\ref{fig:LineConfig} shows time evolution of a linear interface.

\begin{figure*}
\resizebox{0.43\columnwidth}{!}{ \includegraphics{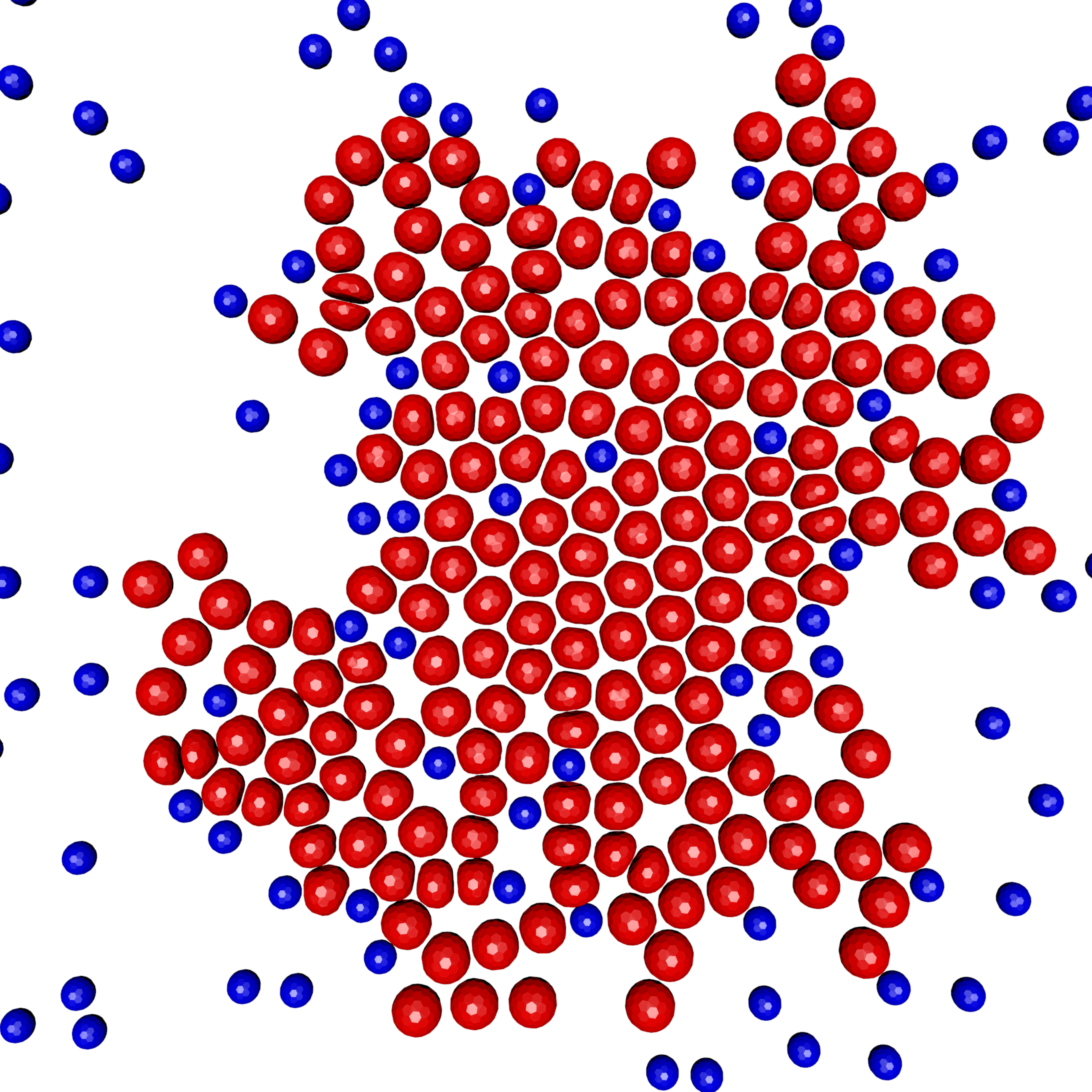} }
\hspace{22pt}\resizebox{0.43\columnwidth}{!}{ \includegraphics{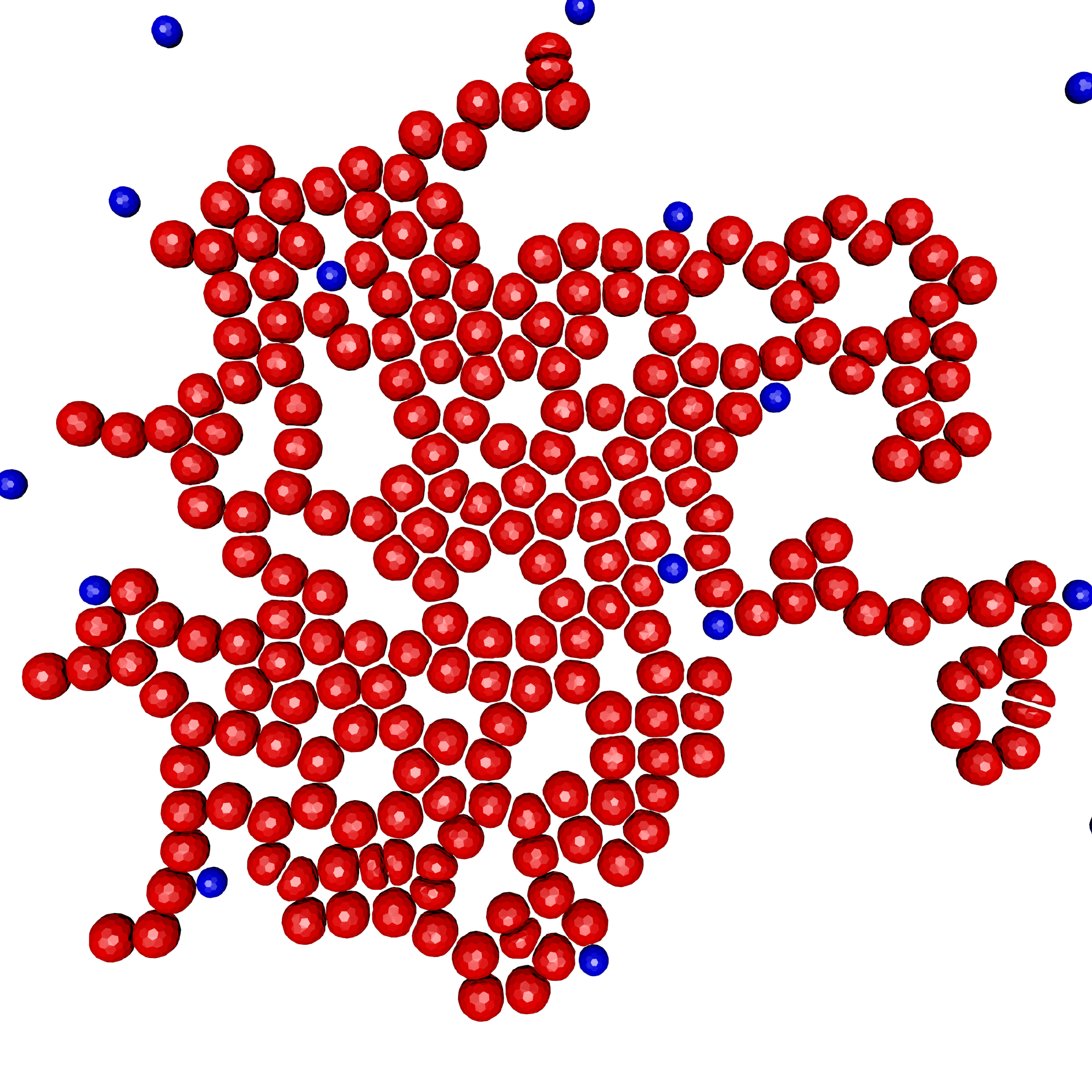} }
\hspace{22pt}\resizebox{0.51\columnwidth}{!}{ \includegraphics{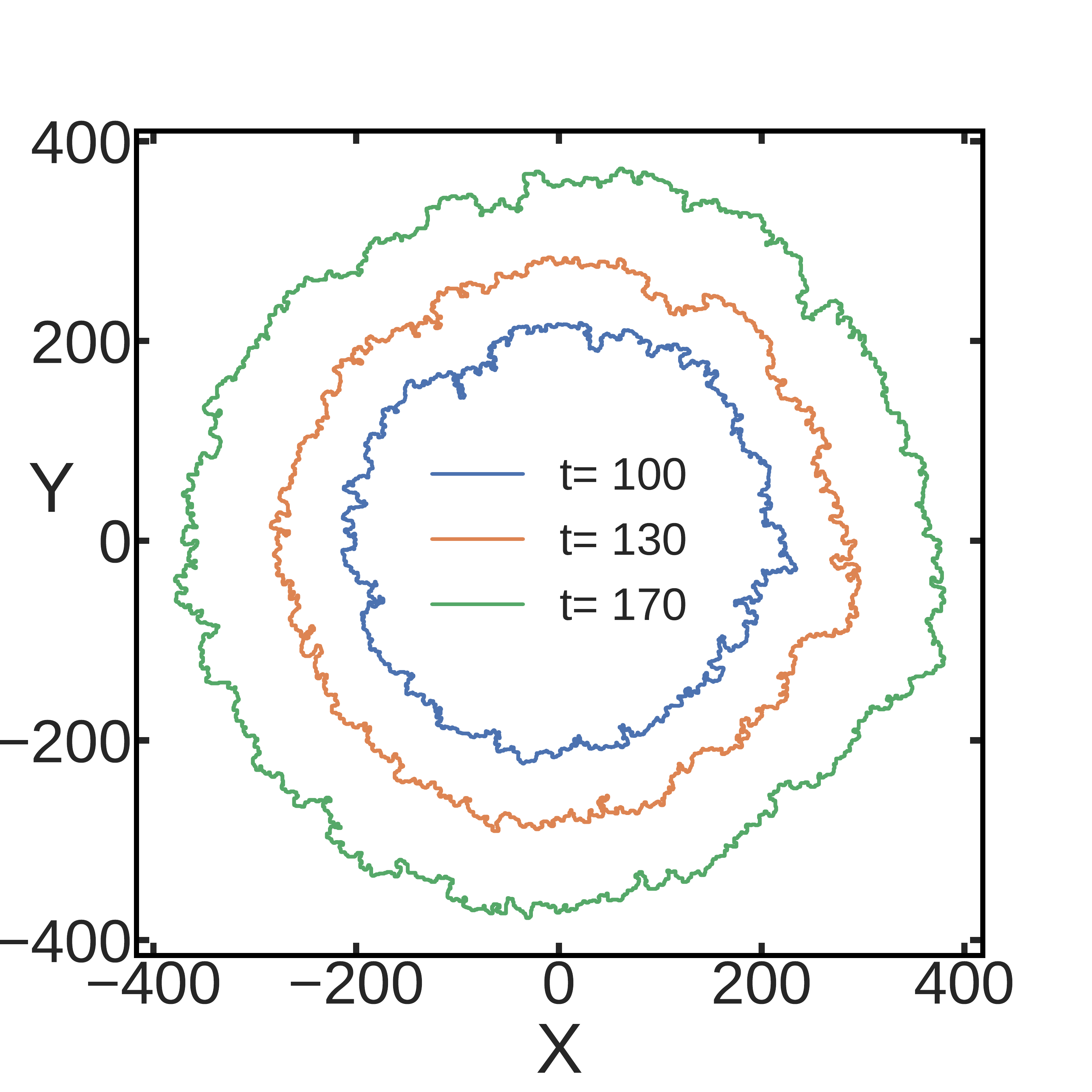} }
\resizebox{0.49\columnwidth}{!}{ \includegraphics{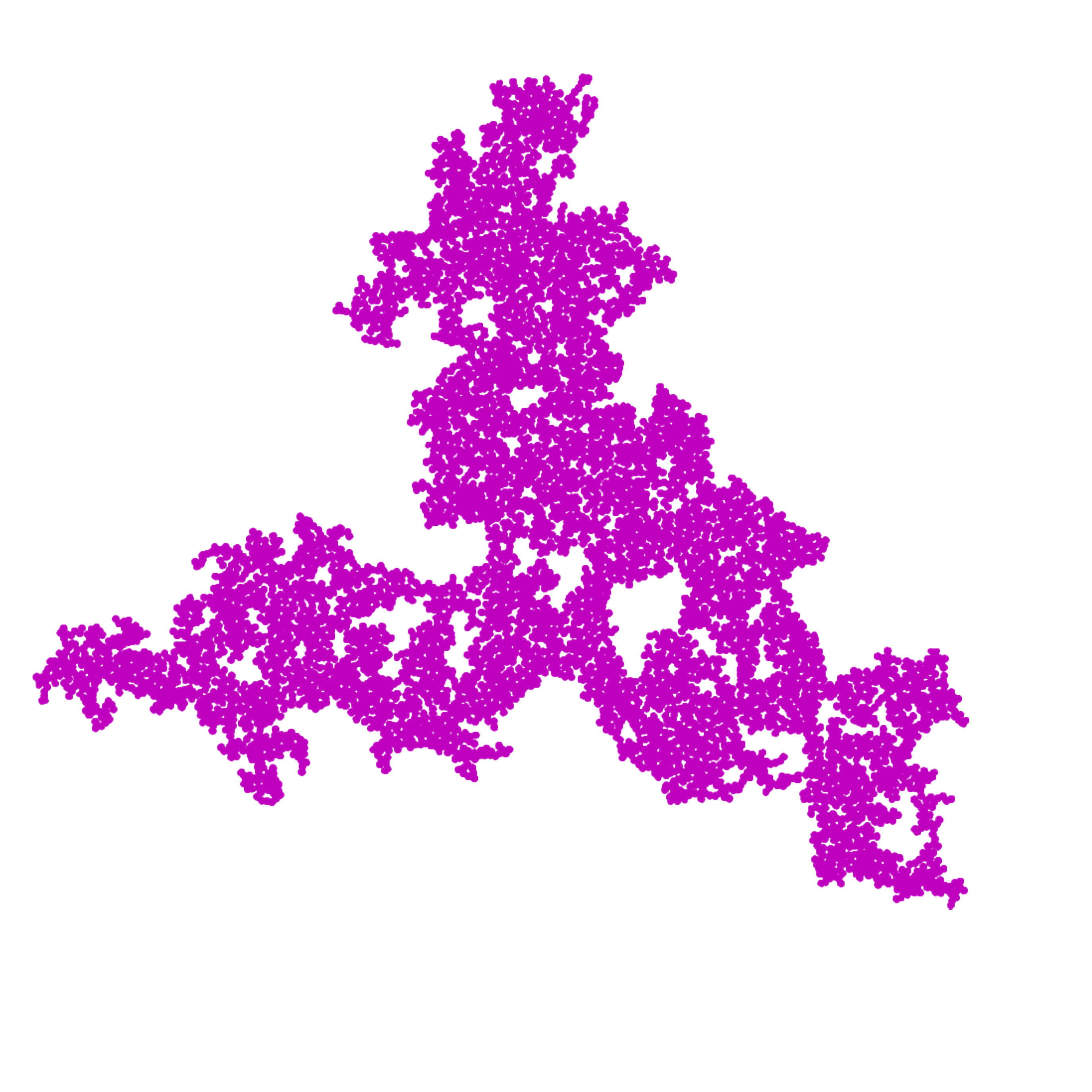} }
\\
a) \hspace{0.5\columnwidth} b) \hspace{0.5\columnwidth} 
c) \hspace{0.5\columnwidth} d) 
\caption{ (a,b) Radially growing colony (red cells) in a medium with quenched disorder (blue cells). (a) At weak and (b) strong cell-cell adhesion strength. (c) Interface evolution at different times at weak cell-cell adhesion strength. The interface has overhangs, but the scaling analysis was done using overhang-corrected interfaces~\cite{Barabasi1995}. (d) Morphology for a system started with a single cell at the centre of a box on a substrate with a high density of quenched disorder at weak cell-cell adhesion strength. Due to the high disorder density, the morphology is not round but instead chiral with branched structures. Eventually the interface becomes pinned by the disorder and the growth stops. The final population of the cell colony
consists of roughly 10,000 cells.
For units, see Table~\ref{tab:SimUnits}.
}
\label{fig:CirConfig}
\end{figure*}

For radial growth, the initial configuration consisted of a single cell at the center of the box of size $800\times800\times1.8$. 
In the low disorder density regime at weak and strong cell-cell adhesion strengths, respectively, $51,200$ and $4,800$ immobile cells were initially distributed at random in the box. In the case of high disorder density at weak cell-cell adhesion strength, the box contained $77,400$ randomly distributed immobile cells, see Table~\ref{tab:DisorderDens}. In the low disorder density regime, the colonies maintained their circular morphology with interface overhangs. In high disorder density regime, however, the cell colonies developed a chiral morphology with branched structures lacking circular interfaces for scaling analysis. Snapshots of circular colony expansion, interface evolution and chiral colony morphology at different times are shown in Fig.~\ref{fig:CirConfig}.

\section{\label{sec:level1}Analyses}

\subsection{\label{sec:level21}Scaling analysis}

\subsubsection{Interface width}

The standard deviation of the front height across a length scale $l$ at time $t$ can be used to define the interface's local width function, $w(l,t)$, which represents the fluctuation around the average height of the interface~\cite{Barabasi1995} as
\begin{equation}
\label{eq:Widthlt}
w(l,t) = \bigg \lbrace \frac{1}{N} \sum_{i=1}^{N} [h_{i}(t) - \langle h_{i} \rangle_{l}]^{2} \bigg \rbrace _{L}^{\frac{1}{2}},
\end{equation} 
where $L$ is the the length of the growing front. 
For radial growth, the height, $h_{i}(t)$, is replaced by the distance $r_{i}(t)$ from the centre of mass of the cell colony. $\langle h_{i} \rangle_{l}$ is the local average of the subsets of arc length $l$, and $\lbrace. \rbrace_{L}$ is the overall average. The fluctuations cannot increase indefinitely, and there exists a saturation time, $t_\mathrm{s}$.

For times greater than the saturation time $t \gg t_{s}$, when the local length $l$ equals the total interface length $L$, the width function $w(L,t)$ represents the interface variance and increases with the interface length $L$ according to $w(L,t) \sim L^{\alpha}$, where $\alpha$ is to referred as the global roughness exponent. For 
times smaller than the saturation time, the interface variance increases as $w(L,t) \sim t^{\beta}$, where $\beta$ is the growth exponent.

For self-affine interfaces the width function $w(L,t)$ satisfies the Family-Vicsek dynamic scaling relation~\cite{Family1985}. This scale invariant behavior implies that the total interface length, $L$, is the only characteristic length scale in the system, and that all length scales are subject to the same physics. However, for $t>t_\mathrm{s}$ the local width function $w(l,t)$ may increase as a function of the local length, $l$, with a local roughness exponent~\cite{Barabasi1995} $\alpha_\mathrm{loc}$ as
\begin{equation}
\label{eq:WidthlLocal}
w(l,t) \sim l^{\alpha_\mathrm{loc}}.
\end{equation} 
The local roughness exponent may differ from the global roughness exponent and can also be derived from the power law behavior of the height-height correlation function, which is defined as
\begin{equation}
\label{eq:CorrFunc}
C(\ell,t) =  \langle \lvert h(x,t) - h(x+\ell,t) \rvert^{2} \rangle_\mathrm{x}  \sim \ell^{\,2
\zeta} \mathrm{\,\,for\,\,} \ell \ll \xi_\parallel,
\end{equation} 
where 
$\xi_\parallel$ is the parallel correlation length of the interface, and $\ell$ is the lateral distance between different points on the interface. 
The the height-height correlation function obeys the scaling ansatz~\cite{Barabasi1995} 
\begin{equation}
\label{eq:FamilyVicCorr}
C(\ell,t) \sim \ell^{\,2\zeta} c(\ell / t^{1/z^{\mathrm{c}}}),
\end{equation}
where $c(x)$ is constant for $x\ll1$ and $c(x) \sim x^{-\,2\zeta}$, for $x\gg1$. In growth models with anomalous behavior, the global roughness ($\alpha$) and dynamic exponents ($z$) calculated from the interface width function differ from $\zeta$ and $z^{\mathrm{c}}$ calculated from the height-height correlation function~\cite{schroeder1993scaling,Sarma1994}. In these models, the scaling function $c(x)$ can be different from constant for $x\ll1$, and the scaling relation for the height-height correlation function becomes
\cite{schroeder1993scaling,kotrla1996nonuniversality}
\begin{equation}
\label{eq:FamilyVicCorrNew}
C(\ell,t) \sim C(1,t) \ell^{\,2\zeta} c(\ell / \xi(t)),
\end{equation}
where $\xi(t) = [t/C(1,t)]^{1/z^{\mathrm{c}}}$. The average step height, $C(1,t)$, grows as
\begin{equation}
\label{eq:C1function}
C(1,t) \sim t^{\,2\lambda}.
\end{equation} 
This modified scaling ansatz, Eq.~\ref{eq:FamilyVicCorrNew}, implies $\alpha = \zeta + \lambda z / 2(1-\lambda)$ and $z = z^{c} / (1-\lambda)$~\cite{kotrla1996nonuniversality}.

For $t \ll t_{s}$ the value of the local width function $w(l,t)$ increases with time with the growth exponent $\beta$ as
\begin{equation}
\label{eq:WidthTime}
w(l,t) \sim t^{\beta}\mathrm{\,\,for\,\,} t\ll t_{s}.
\end{equation} 

\subsubsection{Structure factor \label{sec:Sk}}

The above real-space analysis takes into account all wavelengths, including short ones, which indicates that finite-size effects can be expected.
As a solution,  the power-law behavior of the power spectrum of the height fluctuations where only long-wavelength modes contribute to the scaling behavior should be analyzed. To calculate the structure factor, 
$S(k,t) = \langle \hat{h}(k,t) \hat{h}(-k,t) \rangle$, 
the $k$th Fourier mode
$\hat{h}(k,t)$ needs to be evaluated.

The Family-Vicsek scaling form of the structure factor can be then given as
\begin{equation}
\label{eq:FamilyVicStruct}
S(k,t) = k^{-(2\alpha + 1)} s(kt^{\frac{1}{z}}), \mathrm{\,\,where}
\end{equation}
\begin{equation}
\label{eq:StructScal}
s(u=kt^{\frac{1}{z}}) = 
 \left\{ \begin{array}{ll}
         \mathrm{const} & \mbox{for $u\gg1$};\\
         u^{2\alpha + 1} & \mbox{for $u\ll1$}.\end{array} \right.
\end{equation}
Here, $\alpha$ is the global roughness exponent and $s(u = kt^{\frac{1}{z}})$ the scaling function. 
Systems with different local and global roughness exponents represent what is known as anomalous roughening~\cite{lopez1997superroughening}. This phenomenon has been observed in various growth models~\cite{krug1994turbulent,lopez1996lack,sarma1996scale} and experiments~\cite{yang1994instability,lopez1998anomalous,Bru1998}. Two known types of anomalous roughening are intrinsic anomalous roughening, where $\alpha_{\mathrm{loc}} < 1$ and $\alpha > \alpha_{\mathrm{loc}}$ , and superroughening, where $\alpha > 1$ and $\alpha_{\mathrm{loc}} = 1$~\cite{ramasco2000generic,lopez1997superroughening}. In such systems
the scaling function, $s(u)$, has the general form
\begin{equation}
\label{eq:StructScalAna}
s(u=kt^{\frac{1}{z}}) = 
 \left\{ \begin{array}{ll}
         u^{2(\alpha - \alpha_{\mathrm{s}})} & \mbox{for $u\gg1$};\\
         u^{2\alpha + 1} & \mbox{for $u\ll1$},\end{array} \right.
\end{equation}
where the spectral roughness exponent, $\alpha_{\mathrm{s}}$, is independent from the global roughness exponent. In system with intrinsic anomalous roughening, $\alpha_{\mathrm{s}} = \alpha_{\mathrm{loc}} < 1$, and $\alpha_{\mathrm{s}}$ is different from global roughness exponent, $\alpha$.

\subsection{\label{sec:chi}Chi-squared minimization}

Chi-squared minimization was used to determine the slope and the $y$-intercept of the line that best fits the data. As every data point in our case was measured ten times, there is a standard error $\sigma_{i}$ that can be associated with individual points in the graphs at different times or lengths. The model's prediction is a lin-lin or log-log straight line $f(x) = b + a \, x $ with parameters $a$ and $b$. The Chi-squared function is calculated by summing the squares of the differences between the model's prediction and the observed data $y_i$, then dividing by the data's variance. It is defined as
\begin{equation}
\label{eq:ChiSq}
\tilde{\chi}^2 = \sum_{i=1}^{N_\mathrm{d}} \frac{(y_i - f(x_i; a,b))^{2}}{\sigma_i^{2}},
\end{equation} 
where $N_\mathrm{d}$ is the total number of data points. The optimal values for the model parameters $a$ and $b$ are obtained by minimization of the chi-squared function. Goodness of fit, the p-values, are calculated from the chi-squared probability function $Q({\chi}^2|N_\mathrm{d}-2)$ corresponding to the probability of accepting the null hypothesis of obtaining the same model parameters if the experiment was performed numerous times with identical setup. A p-value near unity indicates that the fit is good, whereas a small p-value indicates that the fit is poor. 

\begin{figure}
\resizebox{0.65\columnwidth}{!}{\includegraphics{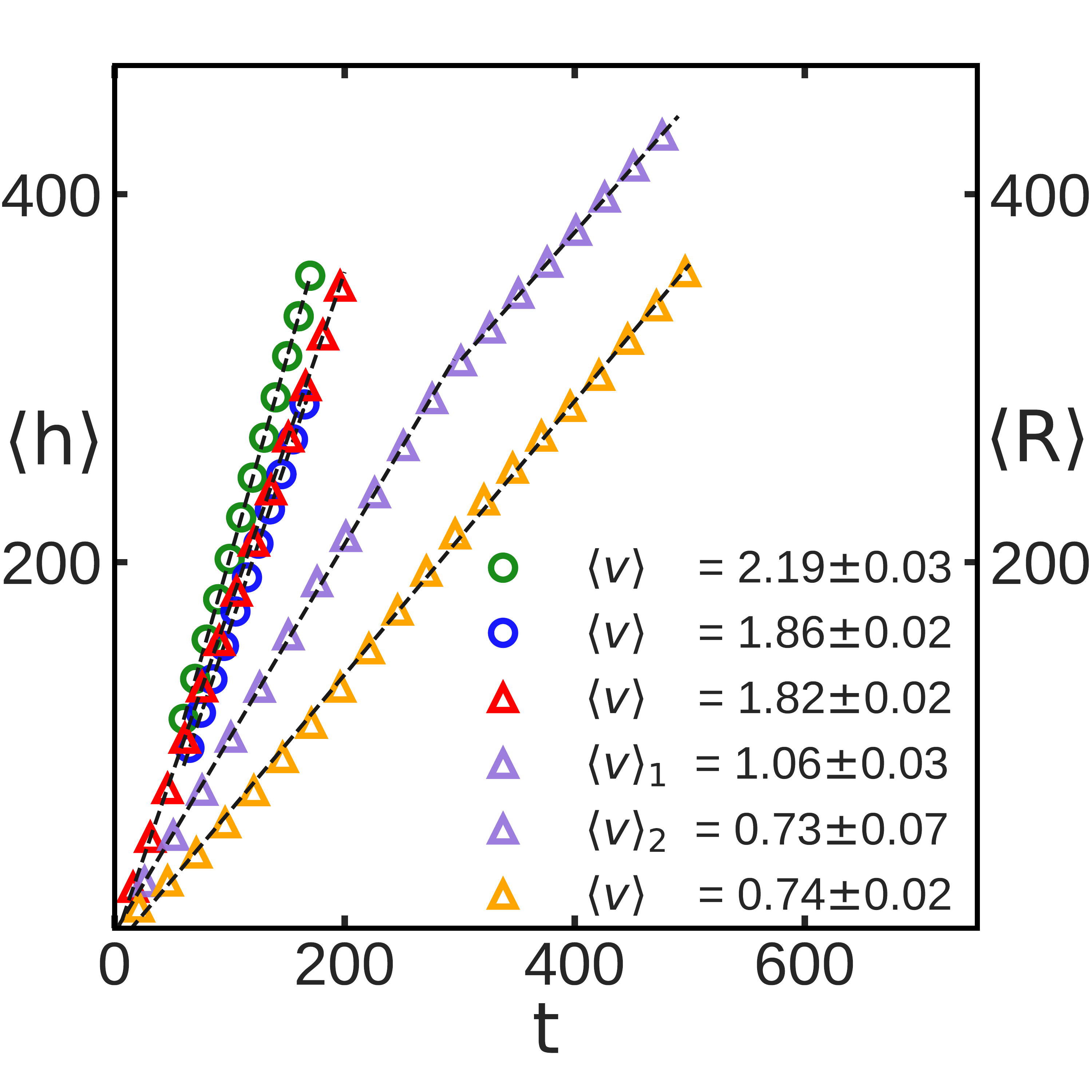}}
\caption{Interface velocity calculated from the time evolution of the mean colony radius ($\langle R \rangle$) and the mean interface height ($\langle h \rangle$) for radially (circles) and linearly (triangles) expanding interfaces, respectively.
For units, see Table~\ref{tab:SimUnits}.}
\label{fig:Velocity}
\end{figure}

\begin{figure}
\resizebox{0.65\columnwidth}{!}{\includegraphics{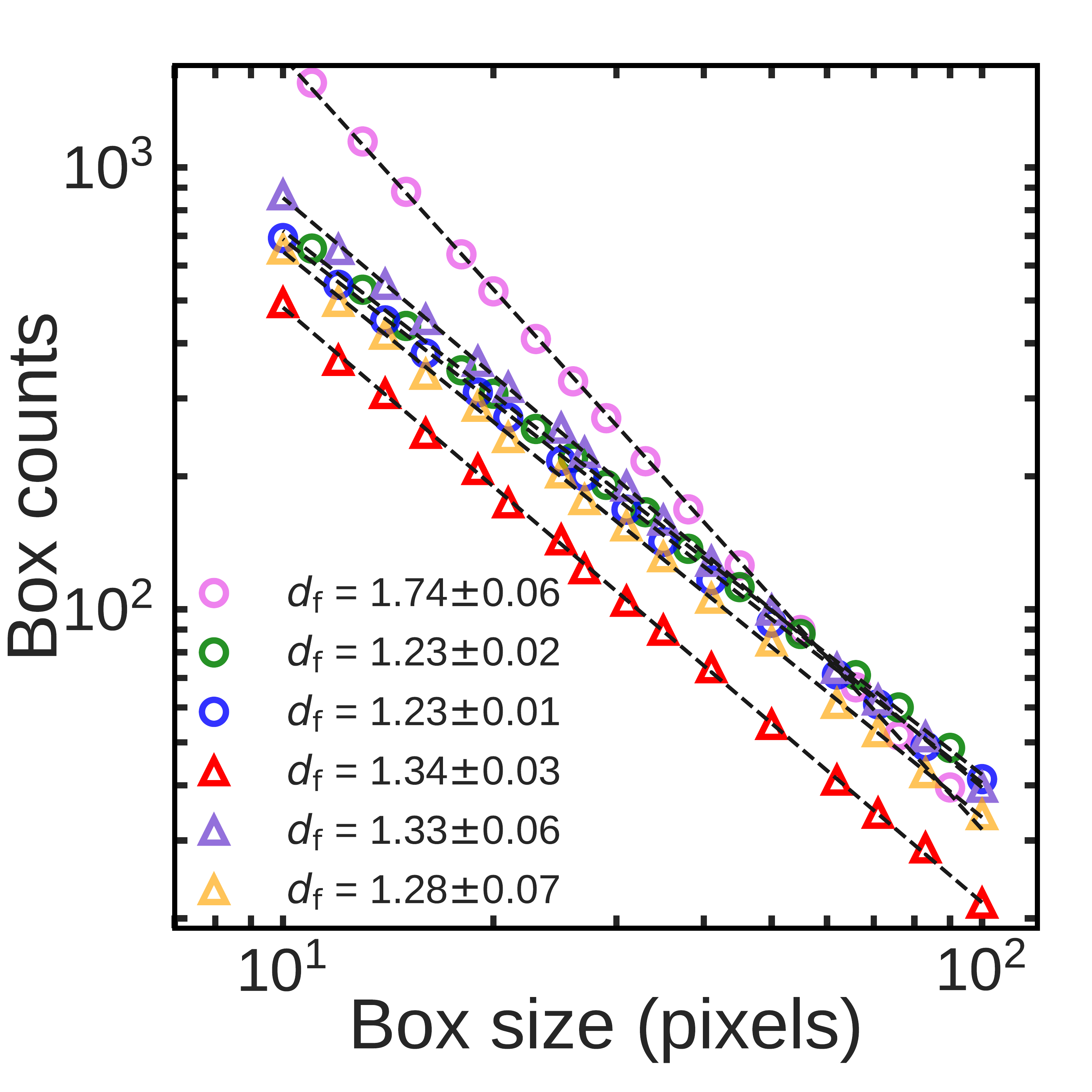}}
\caption{Fractal dimension ($d_\mathrm{f}$) determined by plotting box counts vs. box size for the linear interface (triangles) 1) at weak (red triangles) and 2) at strong (orange triangles) adhesion strength, and 3) at weak cell-cell adhesion in high disorder density  (purple triangle). For the radially expanding interface (circles) 1) at weak (green circles) and 2) strong (blue circles) cell-cell adhesion, 
and high disorder density (violet circles). The colony fractal dimension, ($d_\mathrm{f}^\mathrm{col}$), for colony expansion at weak cell-cell adhesion began with a single cell in the centre of a box on a substrate with a high density of quenched disorder. This system does not have a dense and round morphology, instead it forms slowly to a chiral morphology with branched structure until the colony interface becomes pinned with the disorder on the substrate. The colony fractal dimension is very close to the DLA fractal model, $d_\mathrm{f} = 1.71$. 
}
\label{fig:Frac}
\end{figure}

\section{\label{sec:level11}Results}

\subsubsection{\label{sec:level23} Interface velocity}

For both linear and radial colonies at low disorder densities, the interfaces move at a constant velocity and do not become pinned by disorder,  Figure~\ref{fig:Velocity}.

In the case of linear interface at weak cell-cell adhesion strength, the interface moves at the velocity of $ \langle v \rangle \! = \! 1.82 \pm 0.02 $. At strong cell-cell adhesion strength the velocity drops to  $ \langle v \rangle \! = \! 0.74 \pm 0.02 $.

In radial interface growth,  the velocities are higher, $\langle v \rangle \! = \! 2.19 \pm 0.03 $ at weak cell-cell adhesion strength, and $ \langle v \rangle \! = \! 1.86 \pm 0.02 $ in the case of strong adhesion strength. 

Two cases deserves special attention: First, in linear growth with weak cell-cell adhesion and high disorder density, the growth slows down and
 there is a crossover from  $ \langle v \rangle_{1} \! = \! 1.06 \pm 0.03 $ at short times to   $ \langle v \rangle_{2} \! = \! 0.73 \pm 0.07 $ at late times.
 Second, in the case of weak adhesion strength and high disorder density in circular expansion, the colonies develop a chiral morphology in which the branches proliferate and get pinned over time, preventing the definition of a circular interface and the evaluation of its velocity. 
\subsubsection{\label{sec:level24} Fractal dimension}

The fractal dimensions of interfaces were evaluated using the box-counting method, Fig.~\ref{fig:Frac}. As a general trend, the fractal dimensions of the linear interfaces are slightly larger compared to the circular ones.  In addition,  the fractal dimensions here are slightly higher than those  in the absence of quenched disorder~\cite{mazarei2022silico}. Table~\ref{tab:exponents} lists the fractal dimensions in the current study, and several past experiments and simulations under different conditions.

The one outlier regarding  the fractal dimension is the system that develops chiral morphology, that is, the circularly growing system with high disorder density. The result is $d_\mathrm{f} = 1.74 \pm 0.06$. This value is within the margin of error to computer simulations of the diffusion limited aggregation (DLA) fractal model with $d_\mathrm{f}^\mathrm{DLA} = 1.71$~\cite{Witten1981,meakin1983diffusion}. This chiral morphology has no well-defined interface but rather a branched structure, and it  has been observed, for example, in bacterial growth on agar plates with a low nutrient concentration~\cite{Ben1994,Fujikawa1989,Matsushita1990}.

\subsubsection{\label{sec:level22} Roughness exponents for linear interfaces}

The interface roughness, $w(l,t)$, was evaluated from Eq.~\ref{eq:WidthTime}.
For linear interface growth with quenched disorder, increasing the cell-cell adhesion strength or the disorder density 
resulted in higher growth exponents ($\beta$) than in the absence of disorder, see Ref.~\cite{mazarei2022silico} and Table~\ref{tab:exponents}. The local roughness exponents, $\alpha_\mathrm{loc}$, were obtained from Eq.~\ref{eq:WidthlLocal}. The exponents have the same value at weak adhesion strength both low and high disorder density. At strong adhesion strength and low disorder density, $\alpha_\mathrm{loc}$ increases slightly, Table~\ref{tab:exponents}. These local roughness exponents are also less than what has been obtained from simulations without quenched disorder, see Ref.~\cite{mazarei2022silico} and Table~\ref{tab:exponents}.

\begin{figure}
\resizebox{0.48\columnwidth}{!}{\includegraphics{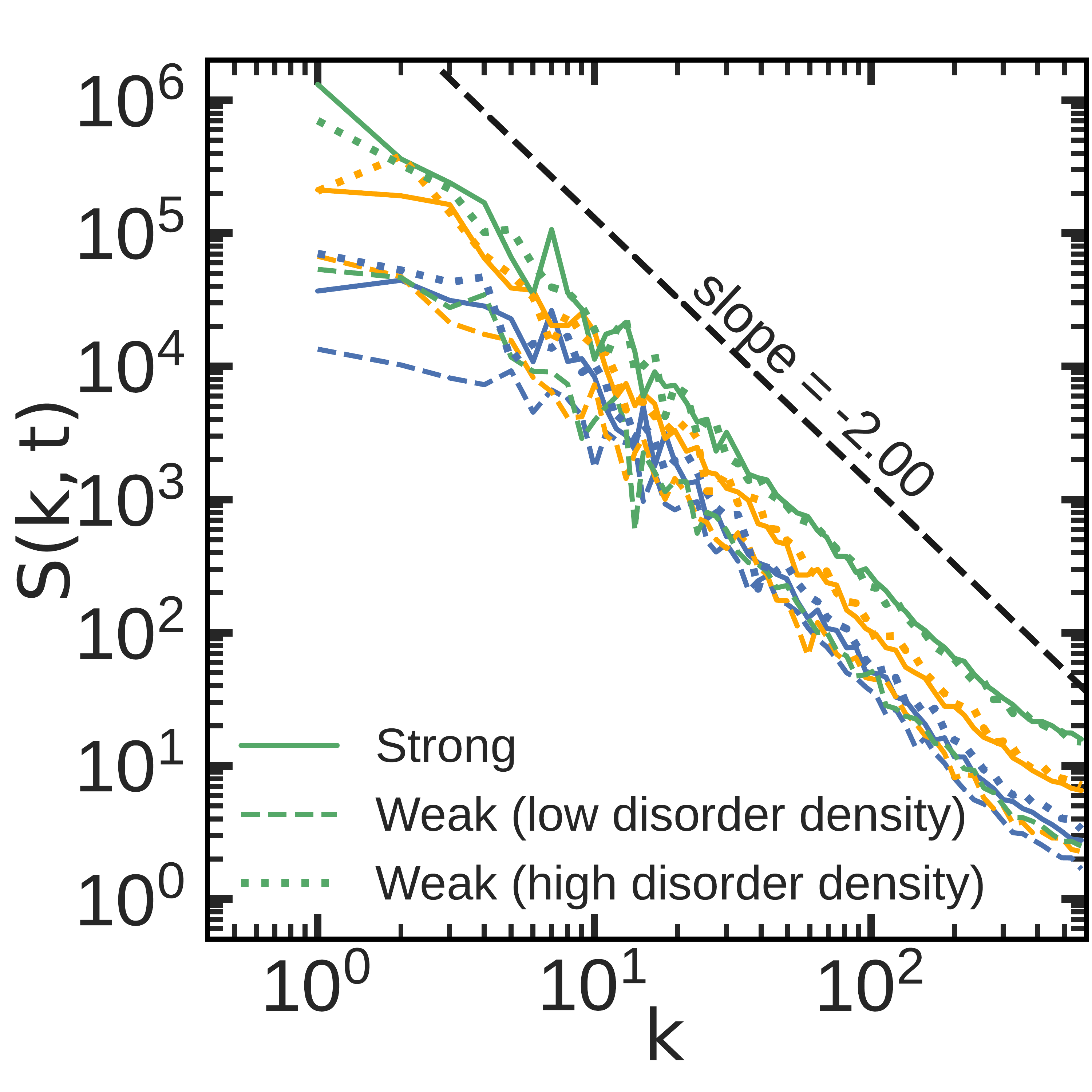}}
\resizebox{0.48\columnwidth}{!}{\includegraphics{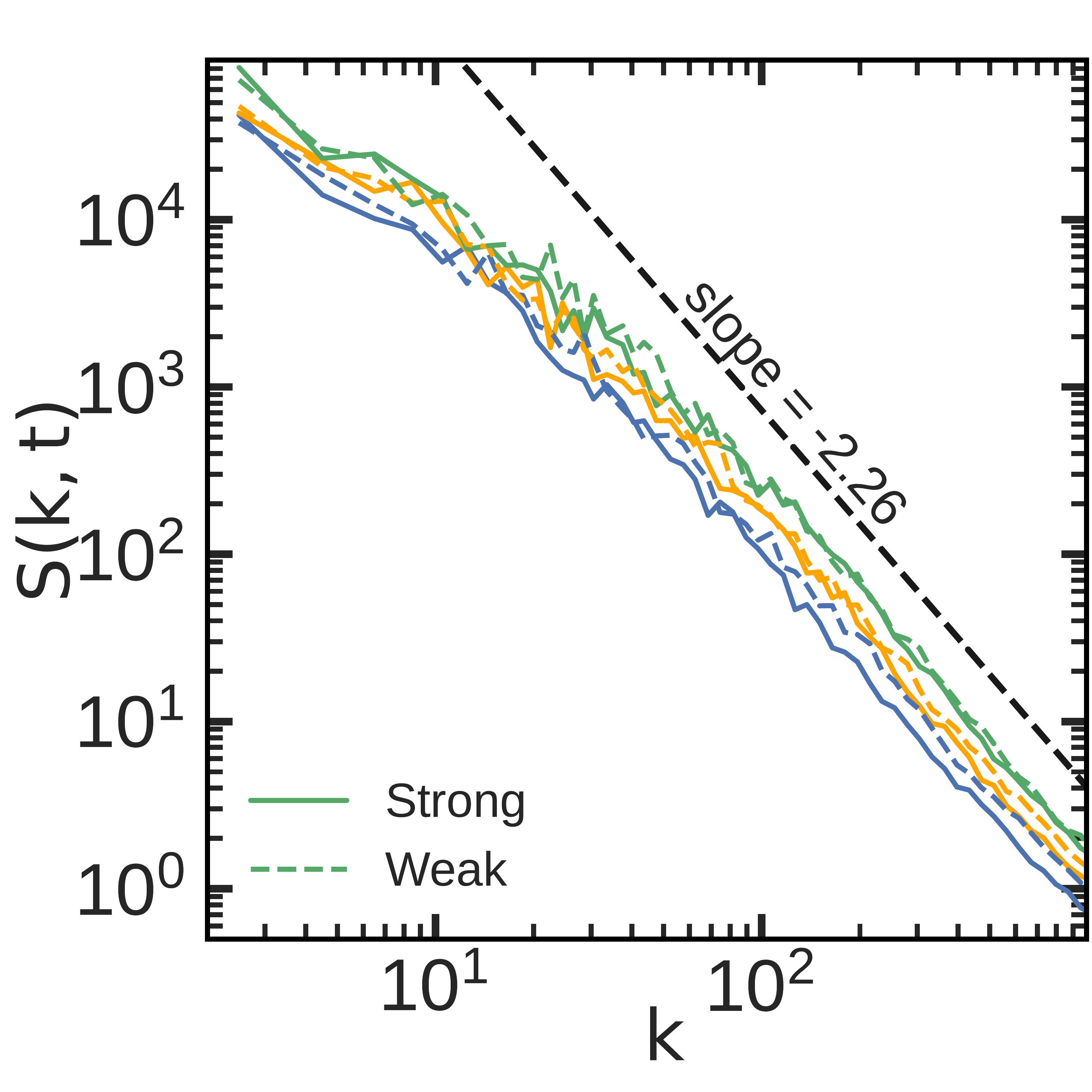}}
\\
a) \hspace{0.4\columnwidth} b) \hspace{0.4\columnwidth} 
\caption{The structure factor (Eq.~\ref{eq:FamilyVicStruct}) measured at three different times, green: long time; orange: intermediate time; blue: short time,
and different conditions (indicated by line type; legend). (a) For the linear interface: Solid lines: strong cell-cell adhesion and low disorder density; dashed lines: weak adhesion, low disorder density; dotted lines: weak adhesion, high disorder density. The black dashed line with a slope of $= -2.0$ is drawn to guide the eye. (b) for the radial interface at low disorder density: Solid lines: strong cell-cell adhesion; dashed lines: weak adhesion. 
The black dashed line with a slope of $= -2.26$ is drawn to guide the eye. The global roughening exponent for each case is reported in Table~\ref{tab:exponents}. For units, see Table~\ref{tab:SimUnits}.
}
\label{fig:StructureFactor}
\end{figure}

The global roughness exponents ($\alpha$) were calculated via structure factor analysis, Eq.~\ref{eq:StructScal}. The results are shown in Fig.~\ref{fig:StructureFactor}. As in the case of $\alpha_\mathrm{loc}$, the global roughness exponents have lower values than those from simulations without quenched disorder~\cite{mazarei2022silico}, see Table~\ref{tab:exponents}. The values are in the same range and independent of the adhesion strength and disorder concentration, whereas for linear colony growth in media without quenched disorder~\cite{mazarei2022silico}, the value of the global roughness exponent depends on the adhesion strength, see Table~\ref{tab:exponents}.

Figure~\ref{fig:Correlation} shows the correlation exponent ($\zeta$) defined via Eq.~\ref{eq:CorrFunc}. Interestingly, the exponent is the same in all cases for linear growth, independent of the disorder density or cell-cell adhesion, Table~\ref{tab:exponents}. The scaling regime, however, increases as cell-cell adhesion increases. 

The correlation exponent was also determined in the absence of disorder based on the data from Ref.~\cite{mazarei2022silico}. In that case, the correlation function shows a crossover between two exponents both at weak and strong cell-cell adhesion,  Table~\ref{tab:exponents}. For shorter scales, the exponents are within the margin error to the value 
$\zeta^\mathrm{weak} \! = \! 0.53$ obtained in the presence of disorder. For longer scales, the exponent crosses over to about $\zeta \approx 0.32$. 

The scaling exponents of linear interface growth at low disorder density at weak adhesion strengths are compatible with KPZ scaling exponents, whereas the global roughness exponent of linear interface growth at weak adhesion strengths in media without quenched disorder is greater than the KPZ global roughness exponent, see Table~\ref{tab:exponents}.
\begin{table*}[!htbp]
    \begin{center}
    \caption{Interface fractal dimension ($d_\mathrm{f}$),  global ($\alpha$) and local ($\alpha_\mathrm{loc}$) roughness exponents, correlation function exponent ($\zeta$; Eq.~\ref{eq:CorrFunc}), growth exponent ($\beta$), 
    and average step height exponent ($\lambda$), in different configurations with different cell-cell adhesion stiffness strengths, and quenched disorder densities, see Tables~\ref{tab:SimUnits} and~\ref{tab:DisorderDens}. For the DLA-like chiral geometry, the fractal dimension is the colony fractal dimension. 
    The exponents for the well-knows cases of KPZ, qKPZ and MBE for one-dimensional interfaces are also given for reference.
    $\dagger$ indicates experiments in heterogeneous media and $\ast$ indicates the crossover with two different regimes.}
    \label{tab:exponents}
	\begin{tabular}{l c c c c c c c}
      \hline
      Configuration & \!\! \!\!\!\!\!\!\!\!\!\!Adhesion 
      &  \!\! \!\!\!\!\!\!\! \! \!\! \!\!\!\!$d_\mathrm{f}$  & $\alpha$ & $\alpha_\mathrm{loc}$ & $\zeta$ & $\beta$ & $\lambda$\\
      \hline
      Kardar-Parisi-Zhang (KPZ)~\cite{Kardar1986}  & - & - & $1/2$ & $1/2$ & - &$1/3$ & - \\
      quenched KPZ  (qKPZ)~\cite{csahok1993dynamics}  & - & - & $3/4$ & $3/4$ & - &$3/5$ & - \\
      Molecular beam epitaxy  (MBE)~\cite{Sarma1994}  & - & - & $3/2$ &
      $1.0$ & - &$3/8$ & - \\
      linear interface at high disorder density  & weak & $1.33 \pm 0.06$ & $0.50 \pm 0.03$ & $0.53 \pm 0.05$ & $0.53 \pm 0.01$ &$ 0.49 \pm 0.07$ & - \\      
      linear interface at low disorder density  & weak & $1.34 \pm 0.03$ & $0.52 \pm 0.04$ & $0.53 \pm 0.02$ & $0.53 \pm 0.01$ &$ 0.33 \pm 0.08$ & - \\
      linear interface  & strong & $1.28 \pm 0.07$ & $0.47 \pm 0.07$ & $0.55 \pm 0.05$ & $0.53 \pm 0.01$ & $ 0.67 \pm 0.07$ & -\\
      circular interface at high disorder density  & weak & $1.74 \pm 0.06$ & - & - & - & - \\
             \,\,\,\, DLA-like chiral geometry &   &   &   &   &   &   & - \\ 
        circular interface at low disorder density  & weak & $1.23 \pm 0.02$ & $0.64 \pm 0.04$ & $0.60 \pm 0.02$ & $0.58 \pm 0.01$ & $ 0.46 \pm 0.13$ & - \\
      circular interface  & strong & $1.23 \pm 0.01$ & $0.63 \pm 0.04$ & $0.62 \pm 0.02$ & $0.58 \pm 0.01$ & $ 0.47 \pm 0.13$ & - \\
      Mazarei \textit{et al.}~\cite{mazarei2022silico} (linear interface)   & weak & $1.22 \pm 0.01$ & $0.75 \pm 0.04$ & $0.59 \pm 0.01$ & $0.51 \pm 0.01^\ast$ &$ 0.28 \pm 0.01$ & $0.02 \pm 0.01$  \\
        &   &   &   &   & $0.31 \pm 0.03^\ast$  &   &  \\ 
      Mazarei \textit{et al.}~\cite{mazarei2022silico} (linear interface)   & strong & $1.26 \pm 0.01$ & $0.52 \pm 0.02$ & $0.62 \pm 0.02$ & $0.55 \pm 0.01^\ast$ & $ 0.25 \pm 0.02$ &  $0.01 \pm 0.01$ \\
        &   &   &   &   & $0.33 \pm 0.04^\ast$  & &  \\ 
      Mazarei \textit{et al.}~\cite{mazarei2022silico} (circular interface)  & weak & $1.13 \pm 0.01$ & $0.95 \pm 0.04$ & $0.66 \pm 0.01$ & $0.59 \pm 0.01^\ast$ & $ 0.40 \pm 0.04$ & $0.37 \pm 0.01$  \\
        &   &   &   &   & $0.32 \pm 0.01^\ast$  & &  \\
      Mazarei \textit{et al.}~\cite{mazarei2022silico} (circular interface) & strong & $1.21 \pm 0.01$ & $0.71 \pm 0.02$ & $0.70 \pm 0.01$ & $0.60 \pm 0.01^\ast$ & $ 0.42 \pm 0.06$ & $0.47 \pm 0.01$  \\
        &   &   &   &   & $0.35 \pm 0.01^\ast$  &  &  \\
      Bru \textit{et al.}~\cite{Bru2003} (circular interface)   & - & $1.12\!-\!1.34 \pm 0.03$ & $1.5 \pm 0.15$ & $0.90 \pm 0.10$ & - & $ 0.38 \pm 0.07$ & - \\
      Huergo \textit{et al.}~\cite{Huergo2011} (circular \& Vero Cells) & - & $1.20 \pm 0.05$ & $0.5 \pm 0.05$ & - & - & $ 0.32 \pm 0.04$ & - \\
      Huergo \textit{et al.}~\cite{Huergo2012} (circular \& HeLa Cells) & - & $1.20 \pm 0.05$ & $0.5 \pm 0.05$ & - & - & $ 0.32 \pm 0.04$ & - \\
      Huergo \textit{et al.}$^\dagger$~\cite{Huergo2014} (linear \& Vero Cells) & - & - & $0.63 \pm 0.03$ & - & - & $ 0.75 \pm 0.05$ & - \\
      Vicsek \textit{et al.}$^\dagger$~\cite{Vicsek1990} (linear interface) & - & - & $0.78 \pm 0.07$ & - & - & - & - \\
      Galeano \textit{et al.}$^\dagger$~\cite{galeano2003dynamical} (circular interface) & - & $1.18 \pm 0.02$  & $ 0.86 \pm 0.04$ & - & - & - & - \\
      Rapin \textit{et al.}$^\dagger$~\cite{rapin2021roughness} (linear interface) & - & -  & - & - & $ 0.58^\ast$ & - &\\
      &   &   &   &   & $0.13-0.25^\ast$  &   & - \\
    \end{tabular}
    \end{center}
\end{table*}

\subsubsection{\label{sec:radial} Roughness exponents for radial growth}

Next, we determine the scaling exponents for radially expanding interfaces. As with linear interfaces, the presence of disorder leads to higher growth exponents ($\beta$) compared to the cases in the absence of disorder, Table~\ref{tab:exponents}. 
Similarly to linear colony growth, the local roughness exponents ($\alpha_\mathrm{loc}$) are in the same range, but somewhat smaller than  without
disorder~\cite{mazarei2022silico}, see Table~\ref{tab:exponents}. 

The global roughness exponents ($\alpha$) 
were calculated via structure factor analysis, Eq.~\ref{eq:StructScal}.
Similar to linearly expanding interfaces, for the radially expanding interface in media with quenched disorder, the global roughness exponents are in the same range and independent of the adhesion strengths, where, as previously mentioned, the global roughness exponent is dependent on the adhesion strength in the absence of quenched disorder, see Fig.~\ref{fig:StructureFactor} and Table~\ref{tab:exponents}. 

The global roughness exponents in media with quenched disorder, similar to linearly expanding interfaces,  are smaller than the global roughness exponents for the radially expanding interface in media without quenched disorder~\cite{mazarei2022silico}, see Table~\ref{tab:exponents}.

\begin{figure*}
\resizebox{0.5\columnwidth}{!}{\includegraphics{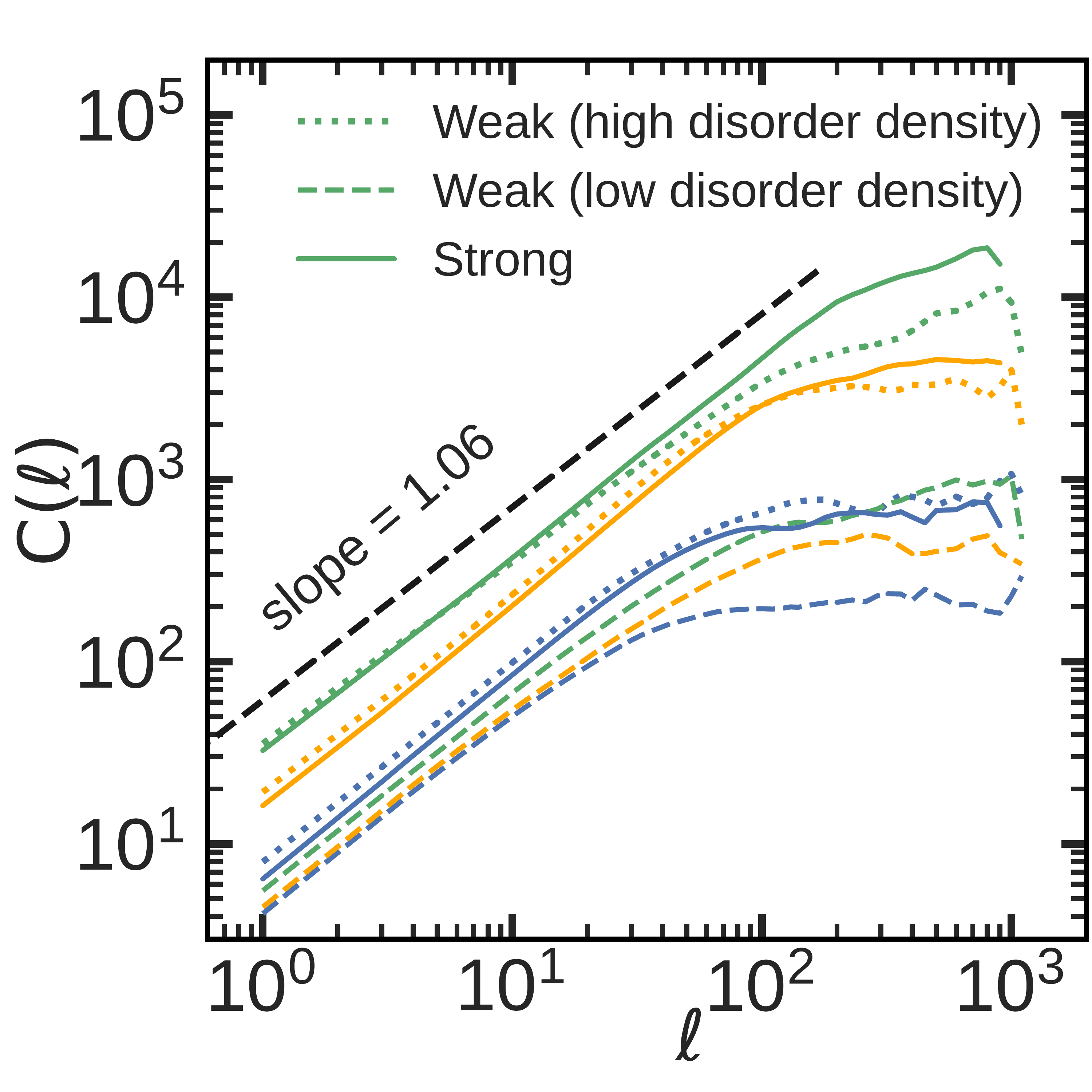}}
\resizebox{0.5\columnwidth}{!}{\includegraphics{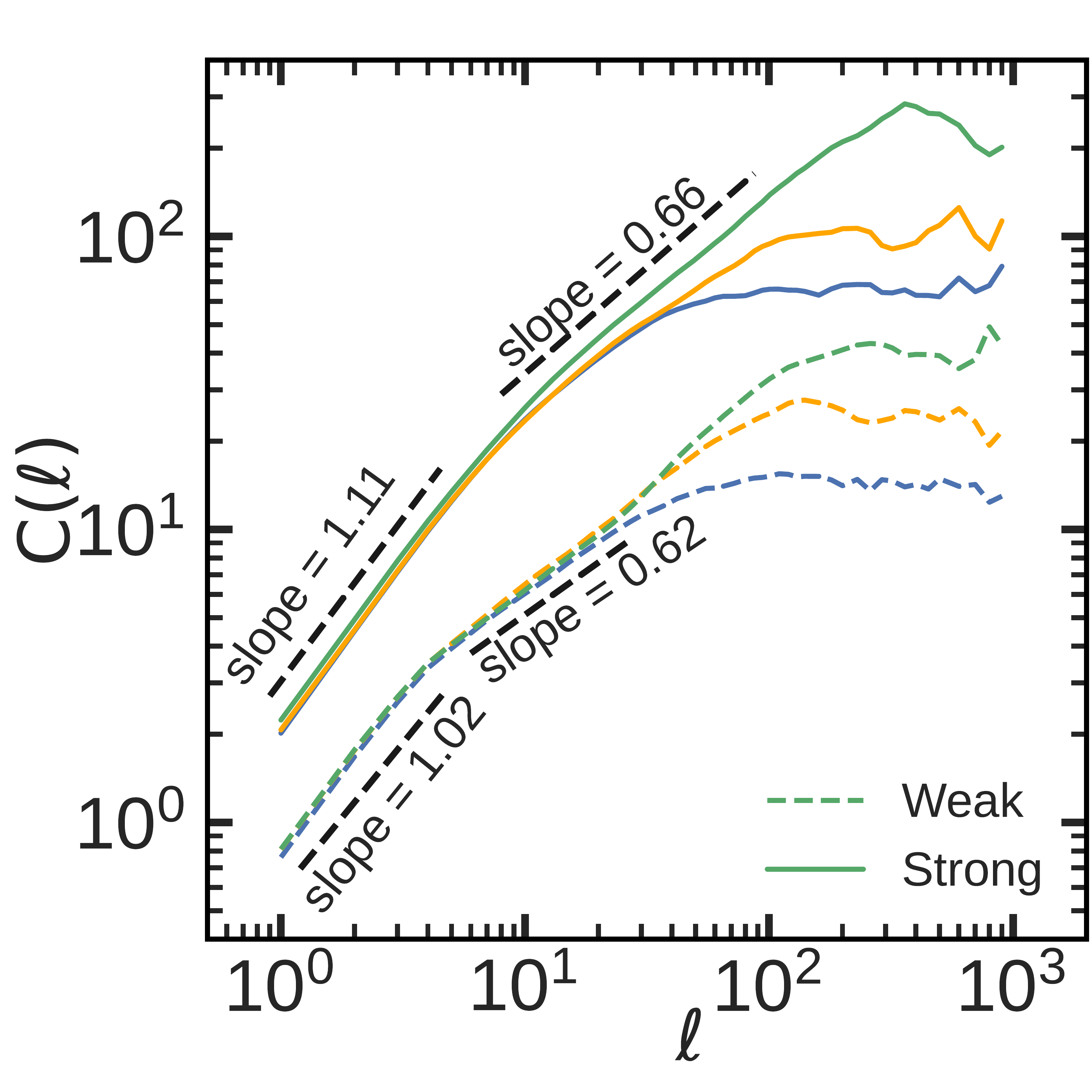}}
\resizebox{0.5\columnwidth}{!}{\includegraphics{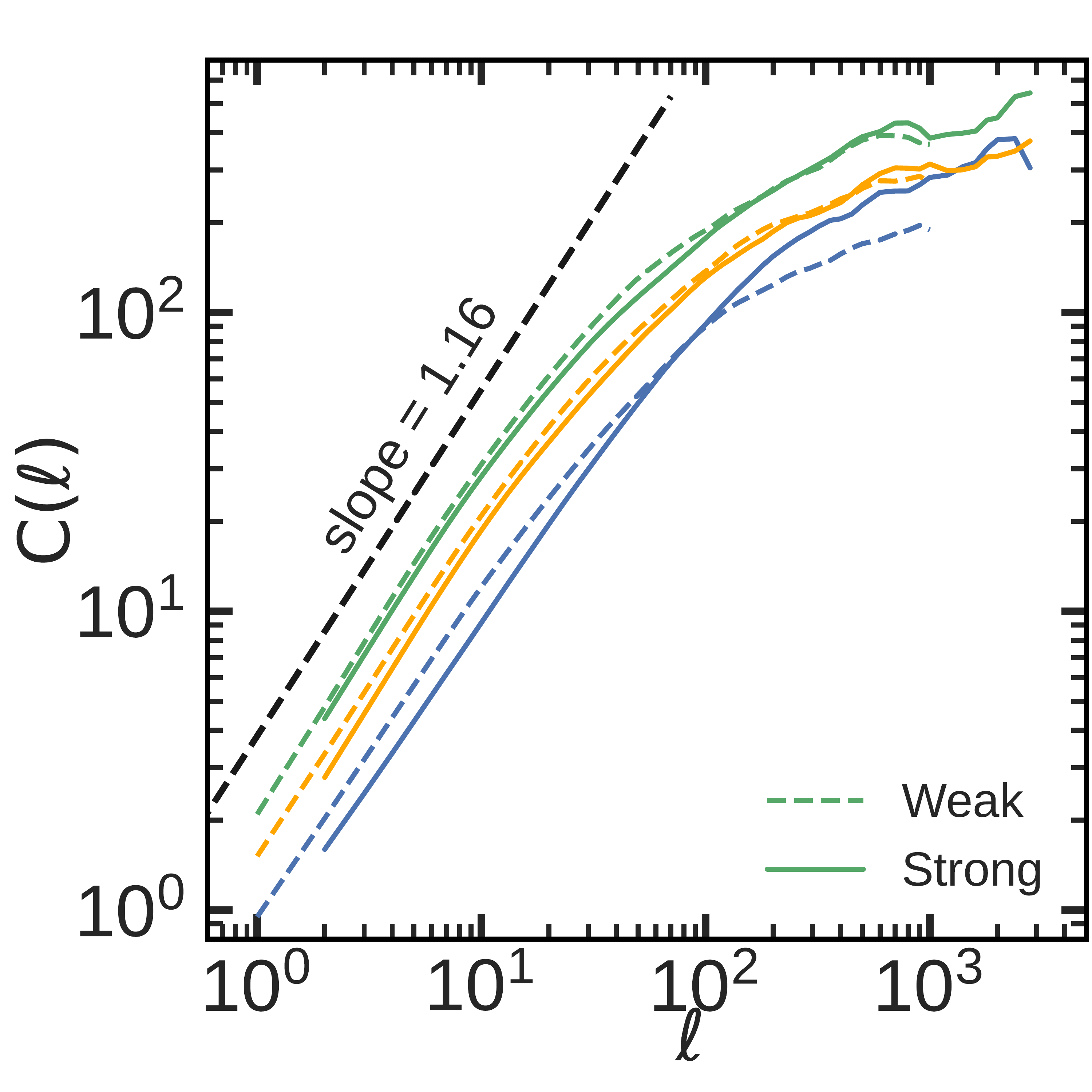}}
\resizebox{0.5\columnwidth}{!}{\includegraphics{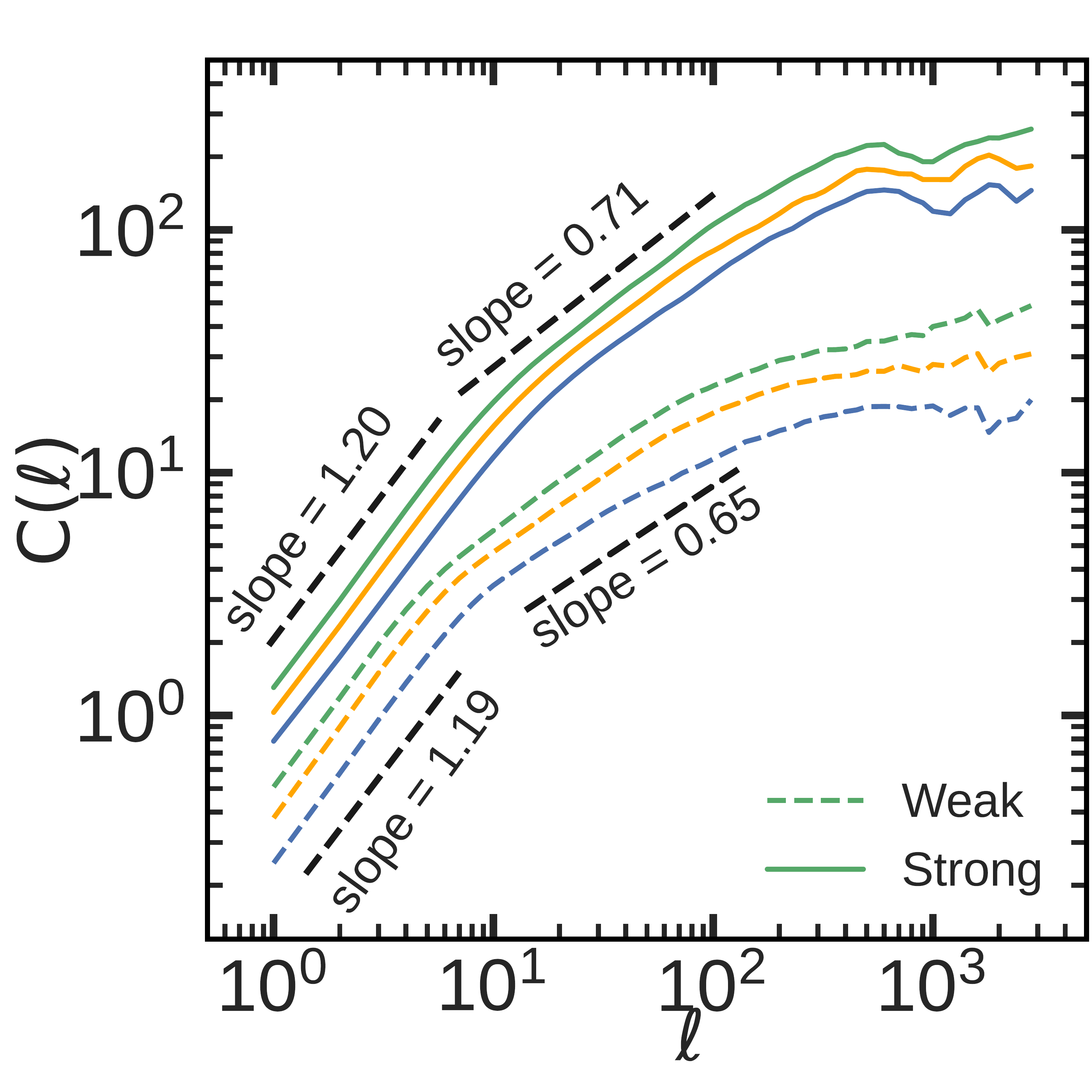}}
\\
a) \hspace{0.5\columnwidth} b) \hspace{0.5\columnwidth} 
c) \hspace{0.5\columnwidth} d) 
\caption{ The height-height correlation function vs length ($\ell$) for the linear interface in a medium (a) with quenched disorder (b) without quenched disorder, and for the radial interface in a medium (c) with quenched disorder (d) without quenched disorder at (green) long (orange) intermediate (blue) short time at different adhesion strengths and disorder densities: Solid lines: strong cell-cell adhesion and low disorder density; dashed lines: weak adhesion, low disorder density; dotted lines: weak adhesion, high disorder density. The correlation function exponents, $\zeta$, are reported in the Table~\ref{tab:exponents}. For units, see Table~\ref{tab:SimUnits}.
}
\label{fig:Correlation}
\end{figure*}

The correlation exponents ($\zeta$) were obtained by determining the height-height correlation function, Eq.~\ref{eq:CorrFunc}, shown in Fig.~\ref{fig:Correlation}, and they have the same value at both strong and weak adhesion strengths, see Table~\ref{tab:exponents}. Figure~\ref{fig:Correlation}
shows the height-height correlation functions for  radially expanding interfaces in media without quenched disorder, and show a crossover with two different correlation exponents for both weak and strong cell-cell adhesion, see Table~\ref{tab:exponents}.

\section{Discussion and conclusions}

Comparison of the present data with previous results for epithelial tissue growth in media without quenched disorder~\cite{mazarei2022silico} shows that quenched disorder can significantly alter the morphology of the interface and cell colony. It also affects cell motility and duplication rate in the colony, resulting in higher fractal dimensions and slower spreading rates. This is consistent with previous experiments for cell colony growth in plain and gel media~\cite{muzzio2016spatio,Huergo2014}.

At the limit of high disorder concentration, colony growth exhibits branched chiral morphologies and the fractal dimension is quite close to the fractal dimension of clusters in diffusion-limited aggregation~\cite{Witten1981}. This has also been observed in bacterial growth on agar plates at low nutrient concentrations~\cite{Ben1994,Fujikawa1989,Matsushita1990}.
In the absence of quenched disorder, increasing adhesion strength
affects the colony morphology and increases the interface fractal dimension~\cite{mazarei2022silico}. 
Here, we have shown that the fractal dimension is independent of the cell-cell adhesion strength for colony expansion on heterogeneous substrates.

In the absence of disorder, adhesion strength is a crucial parameter that generates both KPZ and MBE-like scaling for colony expansion at strong and weak cell-cell adhesion strength, respectively~\cite{mazarei2022silico}. Here, we have demonstrated that in the presence of quenched disorder, the local and global roughness exponent are independent of adhesion strength. This indicates that the effect of adhesion strength on interface roughness and morphology become insignificant on heterogeneous substrates. Disorder does, however,  alter the growth exponent. The growth exponent for linear colony expansion at strong adhesion are within the margin of error of those obtained by Huergo \textit{et al.} in experiments of linear interface expansion of Vero Cells in a gel medium~\cite{Huergo2014}. However, in the case of the linear interface expansion at weak adhesion with both high and low disorder, the growth exponent is different from the one reported by Huergo \textit{et al.}

At low disorder density and weak adhesion, colony expansion from a single line showed KPZ-like scaling. This is in contrast to the situation without disorder~\cite{mazarei2022silico}. 
Although, increasing 
adhesion strength and disorder density does not affect the local and global roughness exponents, the higher disorder density leads to higher growth exponents and makes the scaling behavior of this configuration unclassified. 

The systems with radial growth at both weak and strong adhesion in media with low disorder density do not show any scaling universality class behavior. This is in contrast to the case of  weak adhesion strength in the absence of disorder that displays MBE-like behavior~\cite{mazarei2022silico}. These results indicate that
the concepts of
scaling behavior in characterizing cell colony growth
should be used with caution
due to sensitivity 
to parameters such as disorder concentration and cell-cell adhesion strengths.

The growth exponents for linear and radial interface growths differ for both strong and weak cell-cell adhesion. The fractal dimensions for radial interfaces are lower than the fractal dimensions for linear interfaces, and the local and global roughness exponents are greater for the radial interface than for the linear interface. 
The substrate topologies for linear and radial colony expansions are different. The radial configuration grows on a plane, whereas the linear configuration grows on a cylinder because of the periodicity in one direction. Both the plane and the cylinder have the same Gaussian curvature. However, the first homotopy groups of a plane and a cylinder are different, 
despite the fact that there is no local difference between the two. A continuous contraction to a point is possible for every closed loop in the plane, but only for some closed loops 
on the cylinder.

Independent of adhesion strength and geometries studied here, 
interface growth
in media without quenched disorder does not belong to the superroughening or the intrinsic anomalous roughness subclasses reported in Refs.~\cite{ramasco2000generic,lopez1997superroughening,Bru1998,yang1994instability,krug1994turbulent,lopez1996lack}. 
The average step height exponent, $\lambda$, and the modified scaling ansatz for the height-height correlation function (Eq.~\ref{eq:FamilyVicCorrNew})~\cite{kotrla1996nonuniversality,schroeder1993scaling}, are also not applicable to the type of anomalous behavior in interface growth in media without quenched disorder. 
The results imply the existence of a new type of anomalous behavior, perhaps necessitating a new scaling ansatz for the  interface width scaling relation.

\begin{acknowledgments}
MK thanks the Discovery and Canada Research Chairs Programs
of the Natural Sciences and Engineering Research Council of Canada (NSERC)
for financial support. MM thanks Western University's Science International Engagement Fund (SIEF) for travel support. 
Computational resources were provided by the Finnish Grid and Cloud Infrastructure FGCI, funded by the Academy of Finland, grant 304973.

\end{acknowledgments}


%

\end{document}